\documentclass[final,letterpaper,twoside,12pt]{article}

\usepackage{epsfig,graphicx,psfrag}
\usepackage{epsfig,graphicx,psfrag,bm,amssymb, setspace}

\usepackage{graphicx}% Include figure files
\usepackage{dcolumn}% Align table columns on decimal point
\usepackage{hepunits}
\usepackage{amsmath,amssymb,amsfonts}
\usepackage{mathrsfs}
\usepackage{color}
\usepackage{hyperref}
\usepackage{subfigure}

\newcommand{\be}{\begin{eqnarray}}
\newcommand{\ee}{\end{eqnarray}}

\begin{document}
\baselineskip 0.6cm

\def\simgt{\mathrel{\lower2.5pt\vbox{\lineskip=0pt\baselineskip=0pt
           \hbox{$>$}\hbox{$\sim$}}}}
\def\simlt{\mathrel{\lower2.5pt\vbox{\lineskip=0pt\baselineskip=0pt
           \hbox{$<$}\hbox{$\sim$}}}}
\def\one{\relax{\rm 1\kern-.25em 1}}

\def\lsim{\mathrel{\rlap{\lower4pt\hbox{\hskip0.2pt$\sim$}}
 \raise2pt\hbox{$<$}}}

\def\gsim{\mathrel{\rlap{\lower4pt\hbox{\hskip0.2pt$\sim$}}
 \raise2pt\hbox{$>$}}}

\begin{titlepage}

\begin{flushright}
UMD-PP-012-026
\end{flushright}

\vskip 1.0cm

\begin{center}

{\Large \bf 
Gauging the Way to MFV
}

\vskip 0.6cm

{\large
Gordan Krnjaic $^{a}$ and Daniel Stolarski $^{b,c}$
}

\vskip 0.4cm

{\it $^a$ Perimeter Institute for Theoretical Physics Waterloo, Ontario, N2L 2Y5, Canada} \\
{\it $^2$ Department of Physics and Astronomy, Johns Hopkins University, 
         Baltimore, MD 21218} \\
{\it $^3$ Center for Fundamental Physics, Department of Physics, \\
         University of Maryland, College Park, MD 20742} \\

\vskip 0.8cm

\abstract{We present a UV complete model with a gauged flavor symmetry which approximately
realizes holomorphic Minimal Flavor Violation (MFV) in $R$-parity
violating (RPV) supersymmetry. Previous work has shown that imposing MFV as 
an ansatz easily evades direct constraints and has interesting collider
phenomenology. The model in this work spontaneously breaks the flavor symmetry
and features the minimum ``exotic'' field content needed to cancel
anomalies. The flavor gauge bosons exhibit an inverted hierarchy so
that those associated with the third generation are the lightest. This
allows low energy flavor constraints to be easily satisfied and leaves
open the possibility of flavor gauge bosons accessible at the LHC. The
usual MSSM RPV operators are all forbidden by the new gauge
symmetry, but the model allows a purely ÒexoticÓ operator which
violates both $R$-parity and baryon number. Since the exotic fields mix
with MSSM-like right handed quarks, diagonalizing the full mass matrix
after flavor-breaking transforms this operator into the trilinear
baryon number violating operator $\bar U \bar D\bar D$ with flavor
coefficients all suppressed by three powers of Yukawa couplings. There
is a limit where this model realizes exact MFV; we compute corrections
away from MFV, show that they are under theoretical control, and find
that the model is viable in large regions of parameter space.
}

\end{center}
\end{titlepage}

% Next command reduces length of table of contents
\setcounter{tocdepth}{2}
\singlespacing
\tableofcontents
\singlespacing

%%%%%%%%%%%%%%%%%%%%%%%%%%%%%%%%%%%%%%%%%%%%%%%%%%%%%%%%%%%%%%%%%%%%%%%%%
%%%%%%%%%%%%%%%%%%%%%%%%%%%%%%%%%%%%%%%%%%%%%%%%%%%%%%%%%%%%%%%%%%%%%%%%%
%%%%%%%%%%%%%%%%%%%%%%%%%%%%%%%%%%%%%%%%%%%%%%%%%%%%%%%%%%%%%%%%%%%%%%%%%
%%%%%%%%%%%%%%%%%%%%%%%%%%%%%%%%%%%%%%%%%%%%%%%%%%%%%%%%%%%%%%%%%%%%%%%%%
%											
%												1. Introduction
%
%%%%%%%%%%%%%%%%%%%%%%%%%%%%%%%%%%%%%%%%%%%%%%%%%%%%%%%%%%%%%%%%%%%%%%%%%
%%%%%%%%%%%%%%%%%%%%%%%%%%%%%%%%%%%%%%%%%%%%%%%%%%%%%%%%%%%%%%%%%%%%%%%%%
%%%%%%%%%%%%%%%%%%%%%%%%%%%%%%%%%%%%%%%%%%%%%%%%%%%%%%%%%%%%%%%%%%%%%%%%%
%%%%%%%%%%%%%%%%%%%%%%%%%%%%%%%%%%%%%%%%%%%%%%%%%%%%%%%%%%%%%%%%%%%%%%%%%

\section{Introduction}

Supersymmetry (SUSY) has long been considered an elegant solution to the hierarchy problem~\cite{Martin:1997ns}.  The LHC is now testing whether weak scale SUSY is found in nature, and as yet has found no evidence for superpartners.  Searches for events with multiple jets and large missing energy -- the generic signature of $R$-parity conserving SUSY -- are currently in good agreement with Standard Model (SM) expectations up to the $\sim$ TeV scale~\cite{CMS-SUSY,ATLAS-SUSY} and place strong constraints on the parameter space that tames the hierarchy problem. 

One possible response to this tension is to abandon $R$-parity, and, with it, the missing energy signatures at the LHC. Unfortunately, allowing $R$-parity violation (RPV) dramatically worsens the SUSY flavor problem -- the parameters in the RPV superpotential all contain flavor indices, and generic values of these parameters are badly ruled out by low energy flavor tests~\cite{Barbier:2004ez}. Furthermore, abandoning $R$-parity introduces operators that violate baryon and lepton number, inducing rapid proton decay. 
One solution to this problem is to impose minimal flavor violation (MFV)~\cite{Chivukula:1987py,Hall:1990ac,Ciuchini:1998xy,Buras:2000dm,D'Ambrosio:2002ex,Cirigliano:2005ck} on the MSSM soft Lagrangian and the RPV superpotential~\cite{Nikolidakis:2007fc}.\footnote{Ref.~\cite{Davidson:2006bd} also considers an MFV-like setup where a single lepton number violating RPV operator is the flavor breaking spurion in the lepton sector.} 
MFV is rooted in the observation that the Standard Model (SM) contains a large flavor symmetry $U(3)_{Q} \times U(3)_{U}\times U(3)_{D}\times U(3)_{L}\times U(3)_{E}$
 which is only broken by the Yukawa couplings. Any hypothetical extension of the SM whose flavor violation arises entirely from Yukawa couplings is said to be MFV.  
 Applying this ansatz to SUSY~\cite{Hall:1990ac,Ciuchini:1998xy} forces the $\cal A$-terms to be aligned with the Yukawa matrices, ${\cal Y}$, and forces the soft masses to be proportional to the unit matrix up to corrections of the form ${\cal Y}^\dagger {\cal Y}$. Applying MFV to theories without $R$-parity~\cite{Nikolidakis:2007fc} forces all the RPV couplings to be suppressed by powers of the Yukawa couplings. 

In~\cite{Csaki:2011ge} it was shown that requiring the flavor breaking spurions to couple holomorphically, a feature that would be expected in any realistic SUSY model, drastically reduces the number of allowed RPV operators. In the absence of neutrino masses, only a single baryon number violating (BNV) operator is allowed at the renormalizable level:
\be
W_{RPV} = ({\cal Y}_{U} \bar U)({\cal Y}_{D} \bar D)({\cal Y}_{D} \bar D) ~~~ ,  
\ee  
and it requires three Yukawa insertions making even the largest element quite small. Since all other RPV operators are forbidden in the superpotential, the accidental lepton number symmetry of the SM is preserved up to Kahler corrections 
suppressed by a high power of the UV cutoff, so proton decay is extremely suppressed. Introducing neutrino masses via the seesaw mechanism does not alter the framework's essential features as all additional lepton violating operators are suppressed
by neutrino masses.  
The MFV structure also allows all flavor constraints to be easily satisfied, and this scenario has interesting phenomenology outlined in~\cite{Csaki:2011ge}.

In the SM, MFV is automatically realized since the Yukawa couplings are the only allowed sources of flavor violation. However, in many models of physics beyond the SM, particularly those like the MSSM where many other flavor violating 
structures are allowed, MFV is merely an intriguing hypothesis.  Complete models that implement MFV are scarce. The most prominent is gauge mediated SUSY breaking~\cite{Dine:1981gu,AlvarezGaume:1981wy,Dimopoulos:1982gm,Dine:1994vc,Dine:1995ag}, which is flavor universal at the scale where SUSY 
breaking is mediated, but RG running induces small MFV deviations away from flavor universality. There are few examples of UV complete models of MFV which deviate from flavor universality and also incorporate RPV. 
%To our knowledge, there are no UV complete models of MFV which deviate from flavor universality and also incorporate RPV. 

In this paper, we build a model which has a gauged flavor symmetry
\be
G_F = SU(3)_{Q} \times SU(3)_{U}\times SU(3)_{D}\times SU(3)_{L}\times SU(3)_{E}\times SU(3)_{N}
\ee
and approximately reproduces the RPV MFV framework of~\cite{Csaki:2011ge} in the IR. Here, the final factor in the gauge group comes from introducing a triplet of right handed neutrino fields $\bar N$. We introduce the minimal field content required to cancel the anomalies of the new gauge symmetry giving us a model which is a supersymmetrized version of~\cite{Grinstein:2010ve}. Similar models have been presented in the context of Grand Unified Theories~\cite{Feldmann:2010yp}, and Left-Right models~\cite{Guadagnoli:2011id}, and a supersymmetrized Left-Right model was briefly considered in~\cite{Mohapatra:2012km}.  This theory contains ``flavon'' superfields\footnote{The flavon fields are denoted $Y$, while the SM Yukawa matrices are denoted $\mathcal{Y}$, and, as we note here, $\mathcal{Y}\propto Y^{-1}$.} $Y$ which break the flavor symmetries, but, unlike many other models which Higgs a gauged flavor group~\cite{Barr:1978rv,Wilczek:1978xi,Ong:1978tq,Froggatt:1978nt,Chakrabarti:1979vy,Maehara:1979kf,Davidson:1979wr,Davidson:1979um,Yanagida:1979gs,Davidson:1979nt}, the Yukawa couplings are inversely proportional to the vacuum expectation value (VEV) of the flavon fields. Therefore the flavor gauge fields associated with the first two generations will be much heavier than that of the third generation. This is a realization of the seesaw mechanism in the quark sector which was first introduced  in~\cite{Berezhiani:1983rk,Berezhiani:1983hm,Berezhiani:1990wn} because the heavier the gauge fields become, the lighter the SM quarks are.  This is a valuable feature because the fields which can mediate flavor changing effects in the first two generations are much more constrained, but, in this setup, they will be very heavy. On the other hand, the third generation gauge bosons are poorly constrained and can be lighter, leaving open the possibility of third generation flavor gauge bosons being discovered at the LHC. 
 
\begin{figure}[t]
\vspace*{1.cm}
\begin{center}
{
\unitlength=0.7 pt
%\SetScale{1.}
%\SetWidth{0.9}      % line    size control
\normalsize    %  letter  size control
{} \allowbreak
\includegraphics[width=0.4 \textwidth]{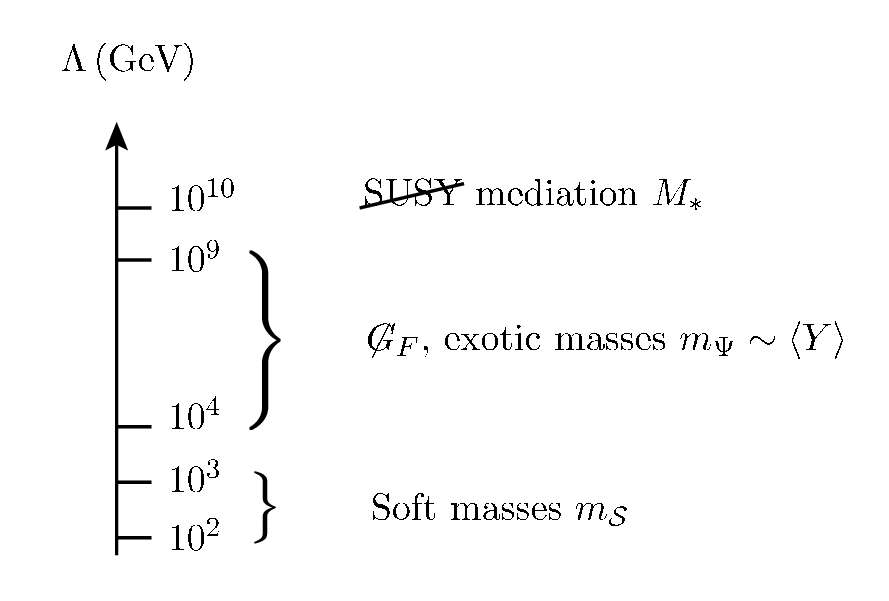}
%
%
%\begin{picture}(-950,-100)(420, 100)
%\Text(-210, 175)[c]{\small $ \Lambda\,(\GeV)$}
%\Line(-150, -20)(-150, 100)
%\ArrowLine(-150, 100)(-150, 101)
%\Line(-150, 80 )(-140, 80)
%\Text(-177 ,120)[c]{\small $10^{10} $   }
%\Line(-80, 80 )(-50,87)
%\Text(-40, 120)[c]{\small SUSY mediation $ M_{*}$  }
%\Line(-150, 65 )(-140, 65)
%\Text(-180,  95)[c]{\small $10^{9} $  }
%\Text(-150, 60)[c]{\Large $\Biggr\}$  }
%\Text(-10, 60)[c]{\small $\displaystyle{\not}{G_{F}}$, exotic masses $m_{\Psi}\sim \langle Y \rangle$  }
%\Line(-150, 17 )(-140, 17)
%\Text(-180, 30)[c]{\small $10^{4} $  }
%\Line(-150, 1 )(-140, 1)
%\Line(-150, -15 )(-140, -15)
%\Text(-150, -9)[c]{\Large $\bigr\}$  }
%\Text(-180, 4)[c]{\small $10^{3} $  }
%\Text(-180, -20)[c]{\small $10^{2} $  }
%\Text(-54,  -10)[c]{\small Soft masses $m_{\cal S} $  }
%\end{picture}
%
}
\end{center}
\vspace*{0cm}
\caption{   The hierarchies of scales in our model. The numerical values given are approximate and can be raised as long as their order remains the same.   }
\label{fig:scales}
\end{figure}

In addition to the MSSM field content, our model has flavon fields which break the flavor gauge symmetry, ``exotics" which cancel anomalies and couple directly to flavons, 
and flavor gauge superfields. The model has the following general structure:
\begin{itemize}
\item SUSY breaking is communicated to the MSSM and exotic fields at a high scale, $M_*$.
\item A gauged flavor group is broken by VEVs of flavons, $\langle Y \rangle$,  at a scale below $M_*$ but above the electroweak and soft scales.
\item The flavon VEVs give mass to the exotic fields at tree level, but they do not directly couple to MSSM-like fields.
\item Mass mixing between exotic and MSSM fields is diagonalized to generate Yukawa couplings with inverse dependence on the flavon VEV.
\item The separation between the flavor and soft scales ensures approximate mass degeneracy among exotics and their superpartners.
\item  $D$-term masses from the flavon VEVs are small compared the hidden sector SUSY breaking contributions to the soft masses.
\item  Exotic interaction eigenstates allow baryon violation while the MSSM-like $\bar U\bar D\bar D$ is forbidden by flavor gauge invariance; the MSSM RPV operator arises when the mass matrix is diagonalized. 
\end{itemize}
An approximate hierarchy of the scales in this model is given in Fig.~\ref{fig:scales}, although the numerical values of the scales are all lower bounds.   We note that many of the details of the model presented here are not necessary to achieve approximate MFV structure as long as the model has the above features. Particular details of our model such as the choice of the flavor group $G_F$ are taken for concreteness, but are not strictly necessary.

%% 

In the limit of $M_* \gg \langle Y \rangle \gg m_\mathcal{S}, v$, MFV is exact, but there are corrections which scale like the ratios of the scales. These corrections come from many different sectors of the model, but most of them are parametrically proportional to the SM Yukawa couplings. Namely, SUSY breaking parameters associated with the first and second generation are much smaller than those associated with the third, but the matrix structure of these parameters is such that they are \textit{not} aligned with the SM Yukawa couplings. Therefore, at leading order, the model presented here is MFV, but there are corrections which realize the ``flavorful SUSY''~\cite{Nomura:2007ap} paradigm.\footnote{RPV in flavorful SUSY was previously considered in~\cite{KerenZur:2012fr} in the context of partial compositeness within SUSY.} Finally, there are corrections coming from $D$-terms of the flavor gauge group which are anarchic in flavor space, but all corrections are under theoretical control and can be sufficiently small to evade all constraints. 

The outline of this paper is as follows.  In section~\ref{sec:model} we describe the quark sector of our model in the SUSY preserving limit. In section~\ref{sec:susy-breaking} we describe how SUSY breaking is communicated and show that flavor violation in the soft Lagrangian inherits approximate MFV structure. We also describe sources of deviation from MFV.  In section~\ref{sec:exp} we consider experimental constraints from direct production and various flavor violating processes, and we show that this model is viable for natural values of the flavor parameters. In section~\ref{sec:leptons} we extend the model to include leptons and we show that MFV leads to a natural realization of pure Dirac neutrino masses. We also briefly consider the more standard seesaw scenario. We give concluding remarks in section~\ref{sec:conclusion}.

%%%%%%%%%%%%%%%%%%%%%%%%%%%%%%%%%%%%%%%%%%%%%%%%%%%%%%%%%%%%%%%%%%%%%%%%%
%%%%%%%%%%%%%%%%%%%%%%%%%%%%%%%%%%%%%%%%%%%%%%%%%%%%%%%%%%%%%%%%%%%%%%%%%
%%%%%%%%%%%%%%%%%%%%%%%%%%%%%%%%%%%%%%%%%%%%%%%%%%%%%%%%%%%%%%%%%%%%%%%%%
%%%%%%%%%%%%%%%%%%%%%%%%%%%%%%%%%%%%%%%%%%%%%%%%%%%%%%%%%%%%%%%%%%%%%%%%%
%											
%											 2. SUSY Model Description
%
%%%%%%%%%%%%%%%%%%%%%%%%%%%%%%%%%%%%%%%%%%%%%%%%%%%%%%%%%%%%%%%%%%%%%%%%%
%%%%%%%%%%%%%%%%%%%%%%%%%%%%%%%%%%%%%%%%%%%%%%%%%%%%%%%%%%%%%%%%%%%%%%%%%
%%%%%%%%%%%%%%%%%%%%%%%%%%%%%%%%%%%%%%%%%%%%%%%%%%%%%%%%%%%%%%%%%%%%%%%%%
%%%%%%%%%%%%%%%%%%%%%%%%%%%%%%%%%%%%%%%%%%%%%%%%%%%%%%%%%%%%%%%%%%%%%%%%%

\section{Gauged Supersymmetric Model}

\label{sec:model}
We begin by describing the quark sector of the model which has a gauged flavor group $G_{F} \equiv SU(3)_{Q} \times SU(3)_{U} \times  SU(3)_{D}$. The full gauge group is $G \equiv G_{SM} \times G_{F}$, with the usual $G_{SM} \equiv SU(3)_{c} \times SU(2)_{L} \times U(1)_{Y}$. In section~\ref{sec:leptons} we 
will extend the model to include leptons, so the flavor group will be enlarged to contain $SU(3)$ factors for the corresponding $L, E$ and $\bar N$ fields. The superfield content of the quark sector is given in Table~\ref{tab:charges}: in addition to the MSSM-like fields, $Q, \bar{u}, \bar{d}$, we add exotic fields $\psi_u, \psi_{u^c}, \psi_d, \psi_{d^c}$ which are also charged under the flavor group as well as SM color and hypercharge; there are no new fields with $SU(2)_L$ charge. 
The $\psi_u\, (\psi_d)$ and $\bar u \,(\bar d\,)$ fields now form a vectorlike pair under the full gauge group, so exotic-SM mass-mixing will be present. 

Finally, the model also features the flavon fields $Y_u$ and $Y_u^c$ which break $SU(3)_Q\times SU(3)_U$, and $Y_d$ and $Y_d^c$ which break $SU(3)_Q\times SU(3)_D$. We now have a supersymmetrized setup of the model presented in~\cite{Grinstein:2010ve}, but here we are forced to add the $Y_u^c$ and $Y_d^c$ superfields to cancel the anomalies that come from the fermions in $Y_u$ and $Y_d$. We will discuss the origin of flavor breaking
 in section~\ref{sec:flavor-vevs}, but we note here that, for the separation of scales we will require, the $Y_{u,d}$ scalar VEVs will be approximately aligned with those of their corresponding  $Y_{u,d}^c$ conjugates.

\begin{table}[t]
\centering
\begin{tabular}{|c|c|c|c|clclcl}
	\hline
		&	$SU(3)_Q$		&	$SU(3)_{U}$		&		$SU(3)_{D}$   &   $SU(3)_{c}$  &   $SU(2)_{L}$ &      $U(1)_{Y}$ \\       
		\hline
	$Q$ 	&	$ \mathbf 3$			&	$ \mathbf 1$			&		  \!\!\! $ \mathbf 1 $                & $\mathbf 3$       & $\quad\mathbf  2$                      & $ +1/6$   \\
	$\overline u$	&	$\mathbf 1$			&	$\mathbf 3$			&		 $\mathbf 1$      & $ \overline{ \mathbf 3}$ &         $\quad \mathbf 1$   &          $-2/3$	\\
	$\overline d$ &    	$\mathbf1$                         &       $\mathbf1$                     &                                           $\mathbf 3$            & $\overline{\mathbf 3}$ &         $\quad \mathbf 1$   &               $+1/3$    \\
	\hline
	\hline
	$ \psi_{u^c}$   &          $ \overline{ \mathbf 3}$  &            $ \mathbf 1$             & $\mathbf    1$  &     $\overline{ \mathbf 3}$ &         \quad $\mathbf1$   & $-2/3$       \\
	$ \psi_{d^c}$  &            $ \overline{ \mathbf 3}$  &        $ \mathbf  1$              &    $\mathbf 1$  &    $\overline{ \mathbf 3}$ &         $\quad \mathbf 1$   &  $+1/3$       \\
	$\psi_{u}$ &            $ \mathbf 1$  &    $\overline{\mathbf 3}$                      &     $\mathbf 1$             & $ \mathbf 3$ &         $\quad \mathbf 1$    &            $+2/3$       \\
	$\psi_{d}$ &            $\mathbf1$  &    $ \mathbf 1$                      &   $ \overline{\mathbf 3}$ &                $\mathbf 3$ &         $\quad \mathbf 1$   &          $-1/3$       \\
	\hline 
	\hline
	$Y_{u}$ &            $\mathbf 3$  &    $\mathbf 3$                      &   $\mathbf  1$   & $\mathbf 1$ &         $\quad \mathbf 1$          &   0       \\
	$ Y^{c}_{u}$ &            $\overline{ \mathbf 3}$  &    $\overline{ \mathbf 3}$                      &    $ \mathbf1$ & $\mathbf1$ &         $\quad \mathbf 1$   &  0       \\
	$Y_{d}$ &            $ \mathbf 3$  &    $\mathbf 1 $                      &   $\mathbf 3$     & $\mathbf 1$ &         $\quad\mathbf 1$     &     0    \\
		$ Y^{c}_{d}$ &            $\overline{ \mathbf 3}$  &    $\mathbf 1 $                      &   $\overline{ \mathbf 3}$ &     $\mathbf 1$ &         $\quad \mathbf 1$ &  0    \\
				\hline 
\end{tabular} \hspace{-0.138cm}\vline
\vspace{0.cm}
\caption{The field content and charge assignments of the quark sector. The top section is MSSM-like fields, the middle section is exotics, and the bottom section is flavons. }
\label{tab:charges}
\end{table}%

The most general, renormalizable superpotential invariant under $G_{SM}\times G_{F}$ is 
\be
W &=& \lambda_{u}  H_{u}  Q  \psi_{u^c}  + \lambda_{u}^{\prime} Y_{u}   \psi_{u} \psi_{u^c} + M_{u}   \psi_{u} \bar u   + W_{Y}  \> + \>(u \rightarrow d )  +   W_{BNV} ~~,
\label{eq:superpotential}
\ee
where the flavon self interactions and baryon violating terms are given by 
\be
 W_{Y} &\equiv& \, \lambda_{Y_{u}} Y_{u} Y_{u} Y_{u} + \, \lambda_{Y_{u}^{c}} Y^{c}_{u} Y^{c}_{u} Y^{c}_{u}    + \mu_{Y_{u}} Y_{u} Y^{c}_{u} 
 ~~, ~~  W_{BNV}  \equiv  \frac{1}{2}\,\eta \, \psi_{u^c}  \psi_{d^c}  \psi_{d^c}   ~~ .
 \label{eq:y-superpotential}
\ee
For the remainder of this paper, whenever down-type interactions have the same structure as their up-type counterparts, they will be omitted for simplicity. 

All dimensionless superpotential couplings ($ \lambda$, $\lambda^{\prime},  \lambda_{Y}, \lambda_{Y^{c}}, \eta$) are {\it flavor universal} 
coefficients and $\mu_{Y}$ is a free parameter.
 Note that $W_{Y}$ contains all allowed $Y_{u,d}, Y^{c}_{u,d}$ interactions and $W_{BNV}$ is the only source of $R$-parity violation. The familiar  
 baryon violating trilinear coupling $\bar u \bar d \bar d$ is forbidden by the flavor symmetry, but will be induced upon spontaneous flavor symmetry breaking (see Sec.~\ref{sec:bnv}).

%%%%%%%%%%%%%%%%%%%%%%%%%%%%%%%%%%%%%%%%%%%%%%%%%%%%%%%%%%%%%%%%%%%%%%%%%
%												2.1 Yukawa Couplings
%%%%%%%%%%%%%%%%%%%%%%%%%%%%%%%%%%%%%%%%%%%%%%%%%%%%%%%%%%%%%%%%%%%%%%%%%

 \subsection{Yukawa Couplings} 
Following \cite{Grinstein:2010ve}, the $Y_{u,d}$ fields  completely break the flavor symmetry prior to electorweak symmetry breaking and give the exotic fields an additional source of mass mixing
\be 
\lambda^{\prime}_{u} \langle Y_{u} \rangle^{\alpha \beta^{\prime}}  ({ \psi_{u^c}})_{ \alpha}  ({\psi_{u}})_{\beta^{\prime}} + M_{u}({\bar u})^{\alpha^{\prime}} ({\psi_{u}})_{\alpha^{\prime}}  \> + \>(u \rightarrow d ) ~~,
\label{eq:superpotential-Yukawas}
\ee
where unprimed and primed greek letters are  $SU(3)_{Q}$ and $SU(3)_{U}$ indices respectively; in our convention, fundamental indices are raised  and anti-fundamental indices are lowered. Following MSSM conventions for Higgs fields, we write the scalar component of $\langle Y_{u}\rangle$ without a tilde; all other scalars will carry tildes. After flavor breaking, $\bar{u}$ and $\psi_{u^c}$ have the same quantum numbers under the remaining unbroken gauge symmetry and can therefore mix with each other. One triplet from the pair matches up with $\psi_{u}$ and gets a Dirac mass that is $\mathcal{O}(\langle Y \rangle )$.  We refer to these exotic mass eigenstates as $\Psi$. The three remaining fields from $\bar{u}$ and $\psi_{u^c}$ are massless before electroweak symmetry and can be identified as the MSSM right handed up superfields $(\bar U, \bar C, \bar T)$. More details of the mass diagonalization are given in the appendix.

\begin{figure}[t]
\vspace*{0.1cm}
\begin{center}
{
\unitlength=0.7 pt
%\SetScale{1.}
%\SetWidth{0.9}      % line    size control
\normalsize    %  letter  size control
{} \allowbreak

\includegraphics[width=0.95 \textwidth]{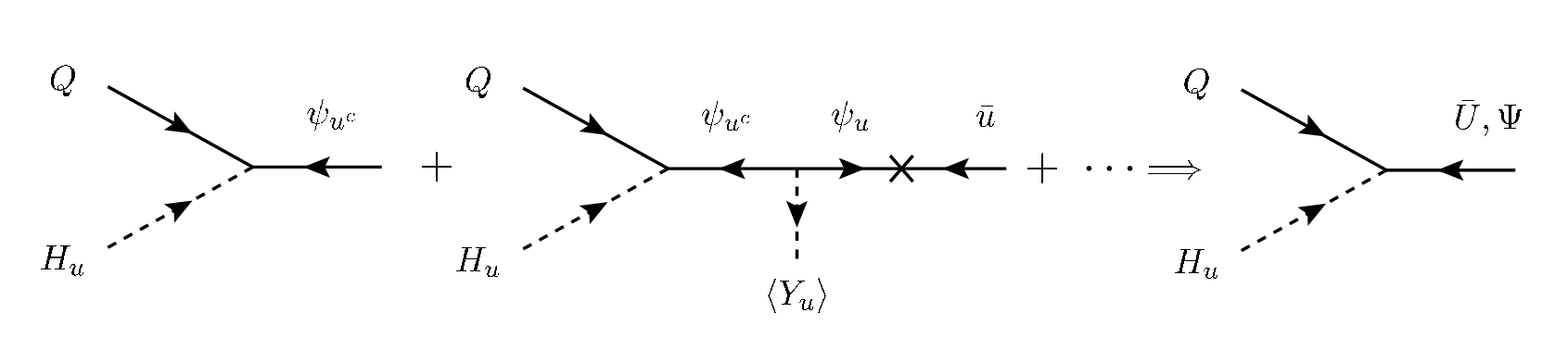}

\parbox{6.7in}{
\caption{   \label{fig:Yukawa}  Diagrams that induce the effective Yukawa couplings in the flavor interaction eigenbasis (left) and the mass
eigenbasis (right). Upon flavor symmetry breaking, the spurion $Y_{u} \to \langle Y_{u}\rangle$ gives dirac masses to $\psi_{u}$ and $ \psi_{u^c}$ -- the former
of which also mixes with $\bar u$ through the supersymmetric mass term $M_{u}\psi_{u} \bar u$. Diagonalizing the exotic masses gives rise to the mass-eigenstate
$\bar U$  which remains massless (prior to EWSB) and is identified with right-handed up-type quark of the MSSM. }
}
}
\end{center}
\end{figure}

Rewriting the Higgs coupling in the superpotential in terms of mass eigenstates gives rise to SM Yukawa couplings, 
\be
 \lambda_{u}  H_{u} Q^{\alpha}  (\psi_{ u^{c}})_{\alpha}  ~~ \longrightarrow ~~
 \lambda_{u} H_{u} Q^{\alpha}       ({\cal V}^{u}_{\alpha \beta^{\prime}}    \bar U^{\beta^{\prime}})  \equiv ({\bf{\cal Y}}_{u})_{\alpha \beta^{\prime}} H_{u} Q^{\alpha} \bar U^{\beta^{\prime}} ~~,
\label{eq:Yukawas}
\ee
where ${\cal{V}}^{u,d}$ is the unitary matrix that transforms interaction states 
into mass eigenstates, and the SM Yukawa matrices are defined by ${\cal Y}_{ u} \equiv \lambda_{u} {\cal V}_{u}$.
 In the $M/\langle Y\rangle \rightarrow 0$ limit, the $\bar u$ are, themselves, the massless eigenstates $\bar U$ 
and do not mix with any exotic fields. Thus, for the lighter generations ${\cal Y} \sim {\cal O}(M/\langle Y\rangle)$ while  
for the ${\cal O}(1)$ top Yukawa coupling, the relationship is more complicated since $M$ and $\langle Y\rangle$ will be comparable.   
Note that the change of basis in Eq.~(\ref{eq:Yukawas}) also induces exotic  $H_{u} Q\Psi$ couplings, which do not exhibit the Yukawa structure. This mass diagonalization is shown schematically in Fig.~\ref{fig:Yukawa}.

Since the Yukawas depend on the ratio $M/\langle Y \rangle$, we demand 
that $\langle Y\rangle \gg M$ for the lighter generations whose corresponding exotic fermions receive large ${\cal O}(\langle Y \rangle)$ masses and 
become exceedingly heavy; for the third generation, the Yukawa coupling is of order unity, so the corresponding exotic partners are naturally lighter. 
Note, however, that {\it the overall scale of $M$ and $Y$ is undetermined}. This feature will become instrumental in suppressing effects that deviate from MFV structure 
after SUSY is broken in section~\ref{sec:susy-breaking}.

%%%%%%%%%%%%%%%%%%%%%%%%%%%%%%%%%%%%%%%%%%%%%%%%%%%%%%%%%%%%%%%%%%%%%%%%%
%												2.2 Trilinear BNV with Exotics  
%%%%%%%%%%%%%%%%%%%%%%%%%%%%%%%%%%%%%%%%%%%%%%%%%%%%%%%%%%%%%%%%%%%%%%%%%

\subsection{Trilinear Baryon Violation}
\label{sec:bnv}

The usual BNV operator in the MSSM is
\be
\lambda^{\prime\prime}_{\alpha\beta\gamma} \bar{U}^\alpha \bar D^\beta \bar{D}^\gamma\,,
\label{eq:MSSM-bnv}
\ee
where $\alpha,\beta,\gamma$ are flavor indices and we have suppressed color indices which are contracted by the epsilon tensor. In this model, that operator is forbidden by the flavor symmetry, and the only source of $R$-parity violation in is given by
\be
W_{BNV} = \frac{1}{2}\,\eta \, \epsilon^{abc} \epsilon^{\alpha \beta \gamma} (\psi_{u^c})_{a \alpha} (\psi_{d^c})_{b\beta } (\psi_{d^c})_{c\gamma}   ~~, 
\ee
 where  $a, b$ and $c$ represent $SU(3)_{c}$ indices. We can rotate from the interaction into the mass basis using the $\mathcal{V}$ matrices and get an operator with characteristic Yukawa suppression 
\be
W_{BNV}  =  \frac{1}{2}\, \eta \, \epsilon^{abc} \epsilon^{\alpha \beta \gamma} ( {\cal V}^{u}_{\alpha \alpha^{\prime}} \bar U_{a}^{\alpha^{\prime}})( {\cal V}^{d}_{\beta \beta^{\prime}} \bar D_{b}^{\beta^{\prime}})( {\cal V}^{d}_{\gamma \gamma^{\prime}} \bar D_{c}^{\gamma^{\prime}}) + {\cal O}(\Psi) ~~.
\label{eq:bnv}
\ee
 The ${\cal O}(\Psi)$ terms are $R$-parity violating couplings between MSSM fields and at least one exotic. These do not play a large role in TeV scale phenomenology because they have both Yukawa suppression as well as suppression from the heavy exotic. 

\begin{figure}[t]
\vspace*{-0.5cm}
\begin{center}
{

\includegraphics[width=0.6 \textwidth]{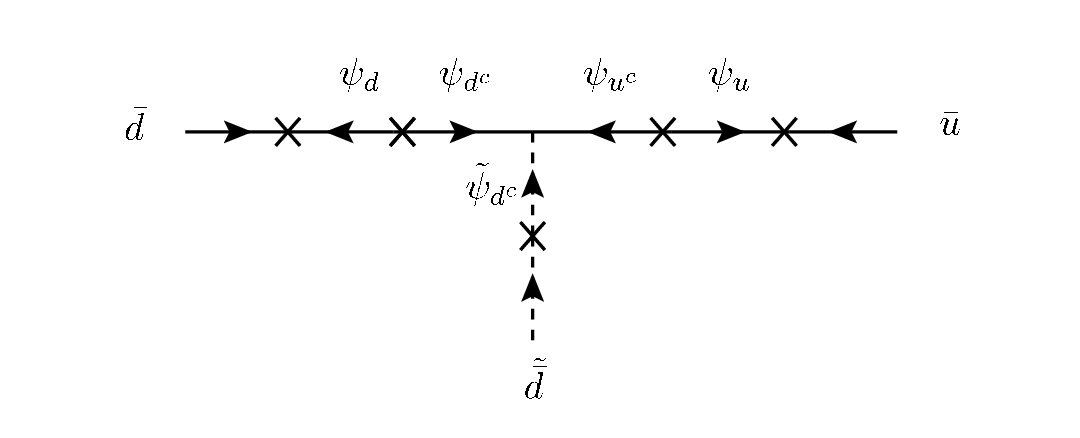}

}
\end{center}
\vspace*{-0.5 cm}
\caption{Example diagram that becomes the trilinear BNV operator $\bar U \bar D  \bar D$ when the exotic interaction eigenstates are diagonalized. Note that the scalar  mass insertion arises
 from the supersymmetric  $F$-term $\lambda_{d}^{\prime \, *} \langle Y_{d}^{*} \rangle M_{d}  \tilde \psi_{d^c}^{*}    \tilde{ \bar d } $ in the scalar potential.  }
\label{fig:bnv}
\end{figure}

From Eq.~(\ref{eq:bnv}), we identify the BNV matrix between MSSM fields defined in Eq.~(\ref{eq:MSSM-bnv}) as 
\be
\lambda^{\prime\prime}_{\alpha^{\prime}\beta^{\prime}\gamma^{\prime}} =  \left(\frac{ \eta }{\lambda_{u}\lambda_{d}^{2} } \right)  \epsilon^{\alpha \beta \gamma} ( {\cal Y}^{u}_{\alpha \alpha^{\prime}})
( {\cal Y}^{d}_{\beta \beta^{\prime}})( {\cal Y}^{d}_{\gamma \gamma^{\prime}}) ~~ ,
\ee
where $(\alpha,\beta,\gamma)$ are $SU(3)_{Q}$ indices and used interchangeably\footnote{Technically, 
the primed indices that accompany $\bar U$ and $\bar D$ do not uniquely correspond to $SU(3)_{U,D}$ indices since the flavor group is now broken and  these are no longer good quantum numbers.} with
$(\alpha^{\prime},\beta^{\prime},\gamma^{\prime})$, and the overall prefactor $ \eta / \lambda_{u}\lambda_{d}^{2} $ is of order one. The way the exotic fields feed into the MSSM BNV operator is shown schematically in Fig.~\ref{fig:bnv}. Note that this is precisely the parametric dependence of the BNV operator in \cite{Csaki:2011ge}, which is the only source of $R$-parity violation
 in the absence of neutrino masses. 

As shown in~\cite{Csaki:2011ge}, the MFV structure of $\lambda^{\prime\prime}$ means that all the entries are very small, with the $tsb$ entry being the largest with size $\sim 1\times10^{- 5} \left( \tan\beta / 10 \right)^2$. Therefore, $R$-parity is a good approximate symmetry, and RPV will only be noticed in very sensitive processes such as the decay of the lightest supersymmetric particle (LSP) and baryon number violation.   The constraints and  phenomenology of these operators are detailed in~\cite{Csaki:2011ge}, and we will review them in section~\ref{sec:exp}.

 %% 

 %%%%%%%%%%%%%%%%%%%%%%%%%%%%%%%%%%%%%%%%%%%%%%%%%%%%%%%%%%%%%%%%%%%%%%%%%
%											2.3	Flavor VEVs & Yukawa Textures
%%%%%%%%%%%%%%%%%%%%%%%%%%%%%%%%%%%%%%%%%%%%%%%%%%%%%%%%%%%%%%%%%%%%%%%%%

\subsection{Flavor Breaking }
\label{sec:flavor-vevs}
Like the non-supersymmetric model in~\cite{Grinstein:2010ve}, our framework does not explain the hierarchies of Yukawa couplings.  Unlike~\cite{Grinstein:2010ve}, this model has the advantage of non-renormalization theorems~\cite{Salam:1974jj,Grisaru:1979wc} for superpotential parameters. Therefore, we accept the tuning in the flavor sector of the SM and get the flavor hierarchy dynamically from superpotential interactions. This means we are trading the technically natural Yukawa couplings for technically natural superpotential parameters. 

To generate nontrivial flavor textures, we can couple our flavons
$Y_{u,d}$ and $Y^{c}_{u,d}$ to a singlet 
\be
W \supset \lambda_{S} S(Y_{u} Y^{c}_{u} - w^{2}) +\frac{1}{2} M_{S} S^{2} + (u \rightarrow d) ~~ , 
\label{eq:ws}
\ee
where $w$ sets the flavor breaking scale and $M_{S}$ is chosen to ensure that $S$ doesn't get a VEV.
The vacuum of this theory is degenerate under all gauge rotations that preserve $\langle Y_u Y^{c}_{u}\rangle = w^{2}$, so minimizing the 
$D$-term potential
\be
\frac{g_{Q}^{2}}{2} \left|  Y_{u}^{*} T^{a}_{\small{ Q}} Y_{u}    -  {Y^{c}_{u}} T^{a}_{\small{ Q}} {Y^{c}_{u}}^{*}   + (u \rightarrow d) + \cdots  \right|^{2}  \> +\> SU(3)_{Q}\to SU(3)_{U,D}  ~~ ,
 \label{eq:dterms3}
\ee
 forces the supersymmetric relation $\langle Y_{u,d}\rangle = \bigl<Y^{c}_{u,d}\bigr>$ up to small corrections that 
arise from the trilinear interactions in Eq.~(\ref{eq:y-superpotential}); we will return to these corrections in section~\ref{sec:squark-masses}. 
 
 Although this mechanism will generate a VEV for $Y$ and $Y^c$,  obtaining the observed Yukawa and CKM matrices requires more structure since the potential derived 
 from Eq.\,(\ref{eq:ws}) doesn't break the full gauge group. Following  \cite{Grinstein:2010ve}, a straightforward generalization involves  
 multiple copies of
 identical $Y, Y^{c},$ and $S$ fields  
 \be
 W \supset   \lambda_{S_{i}}\, S^{i}\left(   C_{ijk}  Y_{u}^{j} (Y^{c}_{u})^{k} - w^{2}_{i} \right) + (u \rightarrow d)  ~~ ,
 \ee
where $C_{ijk}$ is a dimensionless matrix of coefficients whose entries are arbitrary for our present purposes. On general grounds, we are free to use $N$ singlets $S_{i}$ and perform
$M$ gauge transformations to exhaust the $SU(3)_{Q} \times SU(3)_{U}$ flavor symmetry and rotate the $Y_{u}^{j}$ and  $Y_{u}^{c\, k}$ VEVs in flavor space.
 For suitably chosen $N > M$, we will have used up all our gauge freedom, but will still have leftover minimization conditions for which  $Y_{u}^{j}$ and $ Y_{u}^{c\, k}$ can no
 longer be rotated into a particular direction of flavor space. Thus we can, in principle, break the full flavor group and give the flavons realistic VEVs that 
  yield the observed Yukawa and CKM matrices. 
 
Having multiple flavon fields does not change the essential structure of the model because $Y_u$ in Eq.~(\ref{eq:superpotential}) is just replaced with $c_i Y_u^i$, which becomes the effective flavon. 
Although the interaction superpotential in Eq.~(\ref{eq:y-superpotential}) becomes more complicated with more copies of $Y_u$ and $Y_u^c$ --  couplings
acquire new indices: $\lambda_Y \to {\lambda_{Y}}_{ijk} \, ,\, \mu_{Y} \to \mu_{Y_{ij}}$ -- the essential requirements remain qualitatively similar. 
For the remainder of this paper, when we consider flavon self interactions we will ignore these details without loss of essential generality; it should be 
understood that any requirements we impose on these parameters in later sections should apply to the effective couplings that take into account contributions from all 
 index combinations.

In this paper, we do not exhaust all possible flavor-breaking mechanisms.  This discussion serves merely to 
demonstrate that the VEVs required for our scenario can, in principle, be engineered without spoiling any of the 
model's essential features; the singlets introduced here play no other role and we will ignore them for the rest of the 
paper. For a  discussion of alternative flavor breaking patterns and mechanisms, see~\cite{Berezhiani:2000cg,Feldmann:2009dc,Alonso:2011yg}.

 %%%%%%%%%%%%%%%%%%%%%%%%%%%%%%%%%%%%%%%%%%%%%%%%%%%%%%%%%%%%%%%%%%%%%%%%%
%												2.4   Running of SM Gauge Couplings
%%%%%%%%%%%%%%%%%%%%%%%%%%%%%%%%%%%%%%%%%%%%%%%%%%%%%%%%%%%%%%%%%%%%%%%%%

\subsection{Running of SM Gauge Couplings }
\label{sec:running}
Since the exotic states carry color and hypercharge, both beta functions are modified in the UV; for sufficiently light exotics, 
both groups also encounter Landau poles below the Planck scale. 
 However, there is no upper bound on either $\langle Y\rangle$ or the mixing parameters $M$ as long as they are below the scale of SUSY breaking mediation $M_*$ (see section~\ref{sec:susy-breaking}). Therefore, the scales $\langle Y\rangle$ and $M$ can be chosen to decouple all exotic fields and avoid Landau poles while maintaing the appropriate ratios to generate the SM Yukawa matrices. Provided that all lower bounds on flavor VEVs are satisfied (see section~\ref{sec:exp}),
 none of the model's essential features depend on the absolute scale. The scales do, of course, affect the LHC discovery potential for exotic fermions and flavor gauge bosons. 

We compute one loop running of the SM gauge couplings including the new fields given in Table~\ref{tab:charges} as well as the fields in the lepton sector described in
 section~\ref{sec:leptons} and given in Table~\ref{tab:lepton-charges-dirac}.  In the lightest plausible realization (see section~\ref{sec:exp}) with third generation exotic masses at $3\, \TeV$ and all other masses scaled up the appropriate inverse Yukawa couplings, the $SU(3)_{c}$ and $U(1)_{Y}$ gauge couplings diverge around $10^{18}$ and $10^{14} \, \GeV$, respectively. 
If naively interpreted as the scale associated with proton decay, the latter 
could pose marginal problems for our model in the lighter exotic regime, however our flavor symmetry forbids lepton violation (see section~\ref{sec:leptons}), so the high scale physics should
also contain our flavor group and forbid the most dangerous operators even at this scale.

This framework unfortunately spoils the high accuracy of $SU(5)$ gauge coupling unification in the MSSM. It is straightforward to add more matter charged under $SU(2)_L$ and $SU(3)_c$ so that all three hit Landau poles around the same scale to accommodate the possibility of unification. This model contains a larger gauge symmetry than the MSSM, so it is possible that there exist more natural embeddings in larger gauge groups which contain the full flavor group. It is also important to note that the computation of Landau poles is specific to the gauge group chosen in this paper; the mechanisms used to realize MFV are more generic and may
work with other flavor groups. We leave Grand Unification possibilities as well as alternative choices of flavor gauge group to future work.

%%%%%%%%%%%%%%%%%%%%%%%%%%%%%%%%%%%%%%%%%%%%%%%%%%%%%%%%%%%%%%%%%%%%%%%%%%
%%%%%%%%%%%%%%%%%%%%%%%%%%%%%%%%%%%%%%%%%%%%%%%%%%%%%%%%%%%%%%%%%%%%%%%%%%
%%%%%%%%%%%%%%%%%%%%%%%%%%%%%%%%%%%%%%%%%%%%%%%%%%%%%%%%%%%%%%%%%%%%%%%%%%
%%%%%%%%%%%%%%%%%%%%%%%%%%%%%%%%%%%%%%%%%%%%%%%%%%%%%%%%%%%%%%%%%%%%%%%%%%
%%%%%%%%%%%%%%%%%%%%%%%%%%%%%%%%%%%%%%%%%%%%%%%%%%%%%%%%%%%%%%%%%%%%%%%%%%
%%%%%%%%%%%%%%%%%%%%%%%%%%%%%%%%%%%%%%%%%%%%%%%%%%%%%%%%%%%%%%%%%%%%%%%%%%
%
%													3. SUSY Breaking 
%
%%%%%%%%%%%%%%%%%%%%%%%%%%%%%%%%%%%%%%%%%%%%%%%%%%%%%%%%%%%%%%%%%%%%%%%%%%
%%%%%%%%%%%%%%%%%%%%%%%%%%%%%%%%%%%%%%%%%%%%%%%%%%%%%%%%%%%%%%%%%%%%%%%%%%
%%%%%%%%%%%%%%%%%%%%%%%%%%%%%%%%%%%%%%%%%%%%%%%%%%%%%%%%%%%%%%%%%%%%%%%%%%
%%%%%%%%%%%%%%%%%%%%%%%%%%%%%%%%%%%%%%%%%%%%%%%%%%%%%%%%%%%%%%%%%%%%%%%%%%
%%%%%%%%%%%%%%%%%%%%%%%%%%%%%%%%%%%%%%%%%%%%%%%%%%%%%%%%%%%%%%%%%%%%%%%%%%

\section{MFV SUSY Breaking}
\label{sec:susy-breaking}

\noindent Thus far, our discussion has assumed exact supersymmetry. When all $F$ and $D$-terms vanish in the vacuum, 
the exotics and their superpartners have degenerate mass matrices (see the appendix) which are diagonalized by the $\cal V$ matrices that 
define SM Yukawa couplings in Eq.\,(\ref{eq:Yukawas}). In order for MFV to be preserved, the mass matrices for the exotic scalars and fermions need to remain approximately degenerate after SUSY breaking. This will allow soft terms to be diagonalized by the same matrices that yield the Yukawa couplings. For example, taking an $\cal A$-term
\be
 H_{u} \tilde Q \tilde \psi_{u^{c}} \to  H_{u} \tilde Q {\cal V}\tilde{\bar U} \propto {\cal Y}_{u} H_{u} \tilde Q \tilde{\bar U}~~,
 \ee
showing that we get the desired MFV structure. In this section~we will describe all the contributions to SUSY breaking and explain how approximate degeneracy among the exotics is preserved. 

In order for our model to be phenomenologically viable, the dominant source of SUSY breaking for the MSSM-like fields will be via a hidden sector mediated by higher dimensional operators.  There are, of course, many alternative SUSY breaking mechanisms one could consider. A particularly interesting example is ``Flavor Mediation'' \cite{Craig:2012di} in which the flavor gauge-interactions mediate soft masses with non-trivial flavor structure. 
However, flavor mediation with an inverted hierarchy of gauge boson masses yields light first and second generation squarks and a heavy third generation, which both spoils MFV and requires more fine 
tuning to cancel quadratic divergences, so we will not adopt this scenario here.
%However, in that scenario, MFV would be lost, but it could allow so-called ``natural SUSY''~\cite{Cohen:1996vb,Brust:2011tb,Papucci:2011wy} to be realized.

The dynamics introduced in section~\ref{sec:model} induce two additional sources of SUSY breaking: 
\begin{itemize}
\item Nonzero $F_{Y}$-terms that arise from flavon self interactions when $Y$ acquires a VEV {\it and} $\lambda_{Y}, \mu_{Y} \ne 0$. This contribution 
will be highly suppressed in the MSSM sector because the $Y$s  couple only to exotics at tree level. 
\item Nonzero $D$-terms for the flavor gauge groups. Those terms are schematically proportional to $g_F^2(\langle Y \rangle^2 - \langle Y^c \rangle^2)$ where $g_F$ is the gauge coupling for the flavor group. The non-degeneracy between $\langle Y \rangle$ and $\langle Y^c \rangle$ is induced both by nonzero  $\lambda_{Y}, \mu_{Y}$ and by the hidden sector SUSY breaking. 
\end{itemize}
Below, we describe the requirements for these two contributions to be small perturbations, so that SUSY is still approximately preserved in the exotic sector and the scalars are roughly degenerate with the fermions.

%%%%%%%%%%%%%%%%%%%%%%%%%%%%%%%%%%%%%%%%%%%%%%%%%%%%%%%%%%%%%%%%%%%%%%%%%
%												3.1 Hidden Sector Breaking
%%%%%%%%%%%%%%%%%%%%%%%%%%%%%%%%%%%%%%%%%%%%%%%%%%%%%%%%%%%%%%%%%%%%%%%%%

\subsection{Hidden Sector SUSY Breaking}
\label{sec:hidden-susy-breaking}

In order to have sufficiently large SUSY breaking, we take an ansatz of a hidden sector field $X$ with $\langle X \rangle = F \theta^2$, and we also assume that the hidden sector communicates to the visible sector through higher dimensional operators suppressed by a high scale $M_*$. We require $M_* \gg \langle Y \rangle$ so that the flavor symmetry is imposed on all allowed soft terms. As we will see in Sec.~\ref{sec:bnvbounds}, the bound on the gravitino mass requires $M_*$ to be relatively large, $M_* \gsim 10^{10}$ GeV, which is completely consistent with having $M_* \gg \langle Y \rangle$.  

Up to irrelevant ${\cal O}(1)$ variations in the coefficients, the soft Lagrangian contains the following $\cal A$, $\cal B$, and soft mass $m_{\cal S}$ terms
  for MSSM-charged fields at the mediation scale 
\be 
%{\cal L}_{\cal S} \!\!\! &\supset&
 \!\!\!\!&&\!\!\!\!\!\! {\cal A}_{\cal S}  \biggl(  H_{u} \tilde{Q} \tilde{\psi}_{{u}^{c}}  +  Y_{u} \tilde{\psi}_{u} \tilde{\psi}_{u^{c}}  
 +   \tilde{\psi}_{u^{c}} \tilde{\psi}_{ d^{c}} \tilde{\psi}_{ d^{c}} +  c.c. \biggr)  +  {\cal B}_{\cal S}\biggl( \,  H_{u}H_{d}   +  \tilde{\psi}_{u}   \tilde{\bar u}  
    +  c.c. \biggr) 
    \nonumber \\ && \quad \quad \quad \quad
  +\,m_{\cal S}^{2}  \biggl(    \tilde{Q}^{\dagger} \tilde{Q} +
    \tilde{\bar u}^{\dagger}      \tilde{\bar u} + 
    \tilde{\psi}_{u}^{\dagger}      \tilde{\psi}_{u}   +  \tilde{\psi}_{ u^{c}}^{\dagger}      \tilde{\psi}_{u^{c}}  
   \biggr) \>+\> (u \rightarrow d)  ~~~~, ~~~
\label{eq:lag-soft}
\ee
where, for simplicity and without loss of essential generality we assume that all allowed higher dimensional operators of the same order. This gives ${\cal A}_{\cal S} \simeq \sqrt{ {\cal B}_{\cal S}}  \simeq m_{\cal S} \equiv F/M_*$.
 
 All the soft parameters are flavor blind since the SUSY spurion is a $G_{F}$ singlet, but flavor violation will arise through the dynamics
 of exotic fields whose masses arise from flavon VEVs. In addition to requiring that the mediation scale is above the flavor scale, we also require that the soft scale is below the flavor scale, $ \langle Y \rangle \gg m_{\cal S}$. This means that in the exotic sector, SUSY breaking masses will be small perturbations on the dominant masses that come from flavor breaking. 
Therefore, exotic boson diagonalization matrices will be modified by additive amounts proportional to the ratio of SUSY breaking to SUSY preserving contributions to the exotic mass squared matrix: 
 $m_{\cal S}^{2}/\langle Y \rangle^{2}$. Because we know that $\langle Y \rangle \gg m_\mathcal{S}$, the deviations from MFV are under theoretical control.

%%%%%%%%%%%%%%%%%%%%%%%%%%%%%%%%%%%%%%%%%%%%%%%%%%%%%%%%%%%%
%					3.2 		Flavon SUSY breaking
%%%%%%%%%%%%%%%%%%%%%%%%%%%%%%%%%%%%%%%%%%%%%%%%%%%%%%%%%%%%

\subsection{Flavon SUSY Breaking }
\label{sec:flavon-breaking}

The exotic scalar  mass matrix also receives SUSY breaking contributions from flavon self interaction operators after flavor breaking.  
In the absence of a symmetry to forbid the $YY^{c}, YYY$ and $Y^{c}Y^{c}Y^{c}$ terms in the 
superpotential, the flavon VEVs break SUSY through nonzero $F_{Y}$ and $F_{Y^{c}}$ terms.
 This breaking induces exotic $\cal B$-terms of the form
\be
\lambda_{u}^{\prime} \left( \mu_{Y}^{*} \langle Y_{u}^{c}\rangle^{*}    + 3\lambda_{Y}^{*}{\langle Y_{u} \rangle^{\!*}}^{2}          \right)\tilde \psi_{u} \tilde \psi_{u^c}  +  (u \rightarrow d) +  c.c. ~~ , 
\label{eq:mu-operator}
\ee
which contribute to exotic scalar (but not fermion) masses and give rise to non-minimal flavor violation.
To ensure that the corrections in Eq.\,(\ref{eq:mu-operator}) are only perturbations away from mass degeneracy, 
we will need $\mu_{Y} \ll \langle Y \rangle$ and $\lambda_{Y}\ll 1$ when we consider constrains on non-MFV interactions in
 section~\ref{sec:flavorful-bounds}.

%%%%%%%%%%%%%%%%%%%%%%%%%%%%%%%%%%%%%%%%%%%%%%%%%%%%%%%%%%%%
%					3.3 		Exotic Scalar Diagonalization
%%%%%%%%%%%%%%%%%%%%%%%%%%%%%%%%%%%%%%%%%%%%%%%%%%%%%%%%%%%%

\subsection{Exotic Scalar Mass Diagonalization}
\label{sec:scalar-diag}

As discussed above and in more detail in the appendix, in the $m_{\mathcal S}, \lambda_{Y}, \mu_{Y} \rightarrow 0$ limit, the matrix which diagonalizes the exotic scalar mass matrix is the same as the one which diagonalizes the fermion exotic mass matrix, $\mathcal V$. The scalar mass matrix is modified when soft-masses are taken into account. After SUSY breaking, the mass eigenbases for exotic states  is no longer identical for fermions and bosons. These effects  
are proportional to  $m_{\cal S}^{2}/\langle Y\rangle^{2} \ll 1$, which is small by construction since the flavor breaking scale is 
a free parameter in the model; only the ratio $M/\langle Y \rangle$ is constrained by the known Yukawa textures. 
For the lighter generations, there is also the hierarchy $\langle Y \rangle \gg  M$ which ensures that (up to overall order-one coefficients) the 
 the diagonalization matrices have the parametric dependence 
\be
{\mathcal V}_{\rm fermion} = {\mathcal V}_{\rm scalar} \sim \frac{M}{\langle Y \rangle}  ~~, 
\ee
 where $\langle Y \rangle$ is a first or second generation VEV and both matrices are proportional to the Yukawa matrix $\cal Y$. 
 
In a more typical regime, the flavon self couplings are nonzero and SUSY is also broken in the hidden sector, so
the elements of the exotic scalar mass squared matrix become shifted by characteristic amounts $\langle Y \rangle^{2} \to \langle Y \rangle^{2} + m_{\cal S}^{2} + \cdots$ where
 the additional terms all arise from SUSY breaking. This shift is reflected in the bosonic diagonalization matrix by small corrections of the form
 \be
 {\mathcal V}_{\rm scalar} &\sim& \frac{M}{ \left[     \langle Y \rangle^{2} + \mu_{Y} \langle Y \rangle+   \lambda_{Y} \langle Y \rangle^{2}  +  m_{\cal S}^{2}  + \cdots     \right]^{1/2}  }  \nonumber \\\nonumber \\
 &\approx& \frac{M}{\langle Y \rangle} \left[    1 
 +\mathcal{O}\left( \lambda_Y \right)
 +  \mathcal{O}\left(  \frac{\mu_Y}{\langle Y \rangle} \right) +   
  \mathcal{O}\left(  \frac{m_\mathcal{S}^2}{\langle Y \rangle^2} \right)
   \right]   \nonumber \\\nonumber \\
 &=&{\cal V}_{\rm fermion} +  \mathcal{O}\left({\cal Y} \lambda_Y \right) 
+ \mathcal{O}\left( {\cal Y} \frac{\mu_Y}{\langle Y \rangle} \right)
  +     \mathcal{O}\left(  {\cal Y}\frac{m_\mathcal{S}^2}{\langle Y \rangle^2} \right)~~,
 \label{eq:scal-diag}
 \ee
where the leading term is consistent with MFV since ${\cal V}_{\rm fermion}$ is proportional to the full Yukawa matrix ${\cal Y}$, but the subleading terms inside the brackets have anarchic flavor structure.
 Nonetheless these terms always appear in  combination with ${\cal V}_{\rm fermion}$ so they are further Yukawa suppressed. 
 
 Note that in Eq.\,(\ref{eq:scal-diag}) we have omitted order one 
 coefficients and only keep track of the parametric dependence of SUSY breaking terms.
Since this expansion is only valid for the first and second generation elements of $\cal V$, we have also omitted the indices on the flavon VEVs,
which, on account of the inverted hierarchy, are predominantly the largest entries in $\langle Y \rangle$.  
In the appendix we give a more detailed treatment of this 
 index dependence but the order-of-magnitude discussion in this section is sufficient to understand the leading corrections away from degeneracy. 
 In section~\ref{sec:exp-indirect}, these additional terms are subject to flavor constraints from their contributions to MSSM $\cal A$-terms and soft masses; we will see that our rough parameterization here will suffice to set  order of magnitude bounds on SUSY breaking corrections.

%%%%%%%%%%%%%%%%%%%%%%%%%%%%%%%%%%%%%%%%%%%%%%%%%%%%%%%%%%%%
%					3.5		Squark A-terms
%%%%%%%%%%%%%%%%%%%%%%%%%%%%%%%%%%%%%%%%%%%%%%%%%%%%%%%%%%%%

\subsection{Squark $\cal A$-terms}

\begin{figure}[t]
\vspace*{0.cm}
\begin{center}
{
\unitlength=0.7 pt
%\SetScale{1.}
%\SetWidth{0.9}      % line    size control
\normalsize    %  letter  size control
{} \allowbreak

\includegraphics[width=0.95 \textwidth]{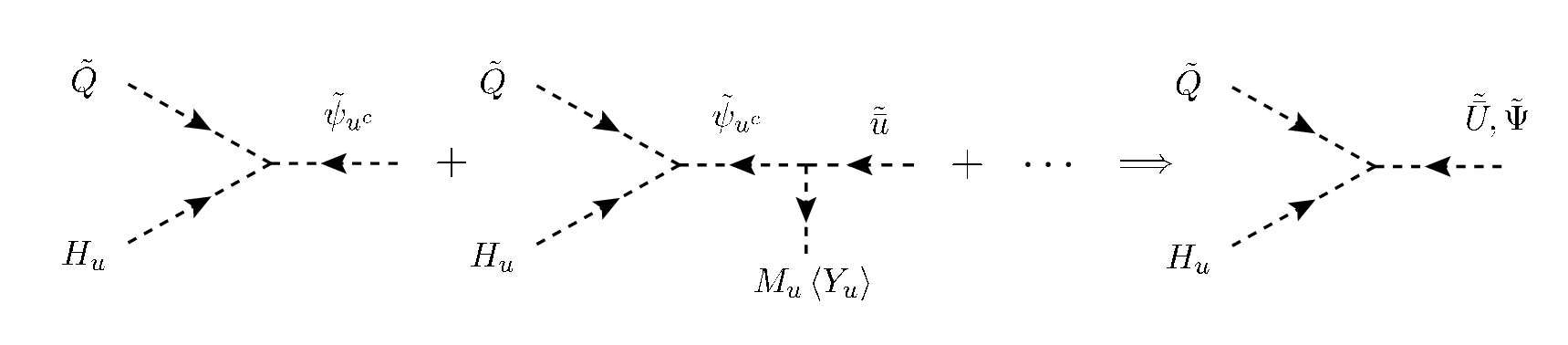}

}
\end{center}
\vspace*{-0.3 cm}
\caption{Interaction basis diagrams that yield $\cal A$-terms (left and middle) and  mass basis diagrams  that connect MSSM fields 
to the mass eigenstates $\tilde{\bar U}$ and $\tilde{\Psi}$ (right). The resulting $H_{u} \tilde Q  \tilde{ \bar{ U}}$ $\cal A$-term is consistent with MFV
up to small corrections that arise from exotic scalar/fermion mass differences once SUSY is broken. }
\label{fig:a-term}
\end{figure}

The $\mathcal A$-terms in the soft Lagrangian of Eq.~(\ref{eq:lag-soft}) can be transformed into the mass eigenbasis upon flavor symmetry-breaking giving the usual MSSM $\mathcal A$-terms,  
\be
{\cal L}_{\cal S} \supset {\cal A}_{\cal S} H_{u} \tilde Q^{\alpha} (\tilde \psi_{u^{c}})_{\alpha} \longrightarrow {\cal A}_{\cal S} H_{u} \tilde Q^{\alpha} ( {\cal V}^{u}_{\alpha \beta^{\prime}}  \tilde{\bar U}^{\beta^{\prime}})  = 
 \frac{{\cal A}_{\cal S}}{\lambda_{u}} ({\cal Y}_{u})_{\alpha \beta^{\prime}} H_{u} \tilde Q^{\alpha} \tilde{ \bar{U}}^{\beta^{\prime}} ~~ .
\ee
which are proportional to Yukawa {\it matrices} and preserve the features of MFV SUSY. This is shown schematically in Fig.~\ref{fig:a-term}, and we see that those diagrams have the same flavor violating couplings as those in Fig.~\ref{fig:Yukawa}. 
Since the mass-eigenvalues/eigenstates for exotic scalars and fermions are identical in the $\langle Y_{u,d} \rangle \gg m_{\cal S}, {\cal A}_{\cal S}, v$ limit (see the appendix), the $\mathcal A$-terms are exactly aligned with the 
Yukawa matrices from Eq.\,(\ref{eq:Yukawas}).  In this limit
the scalars get nearly all their mass from SUSY preserving effects and the MFV structure 
arises in diagrams where the $ {\cal A}_{\cal S}H_{u} \tilde Q \tilde{\psi}_{u^{c}}$  supply the only SUSY breaking  vertices; the rest of the diagram involves  
supersymmetric $\tilde{\psi}_{u}, \tilde{\psi}_{{u}^{c}} $ and $\tilde{\bar u}$ mass-mixing insertions. 

%% 

The differences between Figs.~\ref{fig:Yukawa} and~\ref{fig:a-term} is that instead of exchanging exotic fermions, $\mathcal A$-terms are generated by exotic scalar exchange. So away from the pure MFV limit of $\langle Y_{u,d} \rangle \gg m_{\cal S}, {\cal A}_{\cal S}, v$, we must include the corrections in Eq.~(\ref{eq:scal-diag}) to the scalar mass matrix.  Setting $\lambda_{u} = 1$ for simplicity, the full structure of a typical  MSSM $\cal A$-term is now
\be
{\cal A}_{\cal S} \left[ {\cal Y}   + {\cal  O}\left( {\cal Y} \lambda_{Y} \right)  +  {\cal  O}\left({\cal Y} \frac{\mu_{Y}}{\langle Y \rangle} \right)    + {\cal  O}\left( {\cal Y}\frac{ m_{\cal S}^{2}}{\langle Y \rangle^{2}} \right)  \right] H_{u} \tilde Q \tilde{\bar U} ~~,
\label{eq:a-full}
\ee
where only the first term $\propto {\cal Y}$ in the brackets is exactly MFV. To preserve the features of MFV SUSY, we will demand that these additional terms be small 
and in section~\ref{sec:exp} we will consider the relevant experimental bounds. 

In this framework, as in the MSSM, only the third generation squarks can receive large $\cal A$-terms; all other such terms will be Yukawa suppressed. Thus, it is easy to accommodate light stop or sbottom LSPs with large mixing between the left and right handed states. Other sparticles will have very small mixing because the $\mathcal A$-terms are small.

%%%%%%%%%%%%%%%%%%%%%%%%%%%%%%%%%%%%%%%%%%%%%%%%%%%%%%%%%%%%
%					3.5		Squark Soft Masses
%%%%%%%%%%%%%%%%%%%%%%%%%%%%%%%%%%%%%%%%%%%%%%%%%%%%%%%%%%%%

 \subsection{Squark Soft Masses}
 \label{sec:squark-masses}

In the interaction eigenbasis, the soft Lagrangian in Eq.\,(\ref{eq:lag-soft}) contains the soft masses 
\be
{\cal L}_{\cal S} \supset m_{\cal S}^{2} \left(  \tilde Q^{\dagger}  \tilde Q + \tilde{\bar u}^{\dagger} \tilde{\bar u} + \tilde{\psi}_{u}^{\dagger} \tilde{\psi}_{u} +  \tilde{\psi}_{u^{c}}^{\dagger} \tilde{\psi}_{u^{c}}  \right) ~~.
\label{eq:soft-interaction-states}
\ee
After diagonalizing the exotic states with unitary matrices, these terms are flavor universal at leading order 
\be
m_{\cal S}^{2} \left(  \tilde Q^{\dagger}  \tilde Q + \tilde{\bar U}^{\dagger} \tilde{\bar U} + \tilde{\Psi}^{\dagger} \tilde{\Psi} \right) ~~,
\label{eq:soft-mass-eigenstates}
\ee
but receive corrections from several sources. 

The leading corrections to flavor universality obey the MFV structure and come from two sources. The first arises from the change of basis going from Eq.\,(\ref{eq:soft-interaction-states}) to Eq.\,(\ref{eq:soft-mass-eigenstates}). If the $\mathcal{O}(1)$ coefficient for $\tilde{\bar u}$ is $c_1$ and that for $\tilde{\psi}_{u}$ is $c_2$, then the soft mass correction goes as $(c_1 - c_2) {{\cal Y}}^{\dagger}{\cal Y} $. Both $c_1$ and $c_2$ are expected to be $\mathcal{O}(1)$, but their difference could be $\mathcal{O}(1)$ or it could be parametrically small depending on the physics that mediates SUSY breaking. This effect applies to the partners of right handed fields, $\tilde{\bar U}$ and $\tilde{\bar D}$, but not to the scalar doublet $\tilde{Q}$ because it does not mix with any exotics. 
  
\begin{figure}[t]
\vspace*{0.5cm}
\begin{center}
{
\unitlength=0.7 pt
%\SetScale{1.}
%\SetWidth{0.9}      % line    size control
\normalsize    %  letter  size control
{} \allowbreak
\includegraphics[width=1.\textwidth]{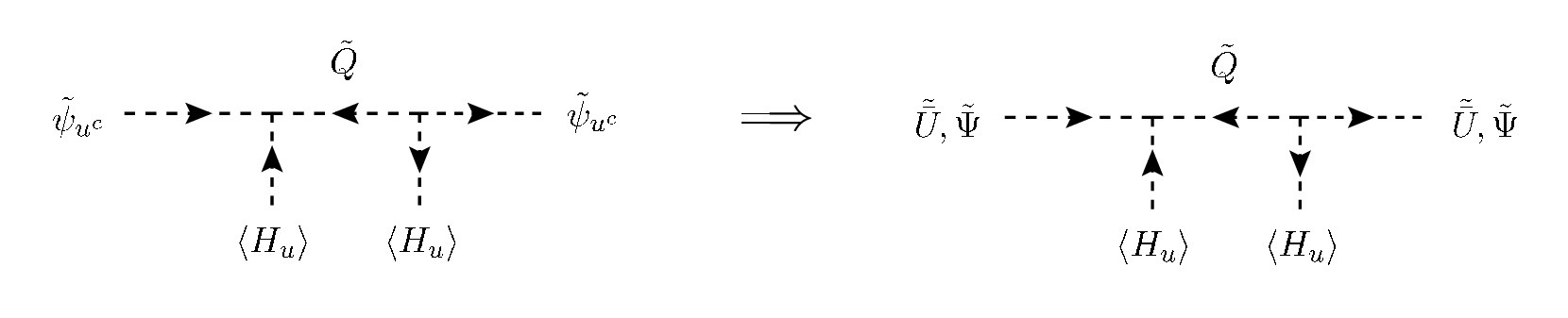}
}
\end{center}
\vspace{-0.2cm}

\caption{ \label{fig:UU-softmass}  Interaction basis (left) correction to $m^{2}_{\cal S} \tilde{\bar U}^{\dagger} \tilde{\bar U}$ and $m^{2}_{\cal S} \tilde{\Psi}^{\dagger} \tilde{\Psi}$ operators; 
mass basis representation of the same diagram (right). Since
the only SUSY breaking interactions in the diagram with  $\tilde{\bar U}^{\dagger} \tilde{\bar U}$ external legs
 arise from ${\cal A}_{\cal S} H_{u} \tilde Q \tilde\psi_{u^{c}}$ vertices (in the interaction basis), the flavor violation in these
processes will be proportional to Yukawa matrices and consistent with MFV. }

\end{figure}

The second source of MFV soft masses are shown in Fig.~\ref{fig:UU-softmass}. The only SUSY breaking vertices in these diagrams are due to the ${\cal A}_{\cal S} H_{u} \tilde Q \tilde \psi_{u^{c}}$  operator which carries Yukawa structure and is brought into the mass eigenbasis by the same matrices that define the Yukawa couplings. This correction to the soft mass exists in the MSSM as long as $\mathcal A$-terms are non-zero and it has the same structure. In both cases it is suppressed by $v^2/m_{\mathcal S}^2$ because it requires two Higgs insertions and the exchange of a heavy scalar.

There are also corrections to the soft mass which are not MFV but still Yukawa-like. These come from the corrections to the matrix that diagonalizes the exotic scalars ${\cal V}_{ \rm scalar} = {\cal V}_{\rm fermion} + \epsilon$, where $\epsilon$
is given in Eq.~(\ref{eq:scal-diag}) and ${\cal V}_{\rm fermion} \propto {\cal Y}$.  This correction depends linearly on the Yukawa couplings and arises because diagrams that induce corrections 
to chirality-preserving soft-masses require at least two insertions of interaction-eigenstate vertices (see Fig.~\ref{fig:UU-softmass}). Thus, transforming to the mass basis involves two diagonalization matrices in the combination  
\be
  \hspace{-0.8cm}({\cal V}_{\rm scalar})^{\dagger}({\cal V}_{\rm scalar}) =({\cal V}_{\rm fermion} + \epsilon)^{\dagger}({\cal V}_{\rm fermion} + \epsilon) 
  = {\cal Y}^{\dagger}{\cal Y} + {\cal Y}^{\dagger}  \epsilon +  \epsilon^{\dagger} {\cal Y} + {\cal O}(\epsilon^{2}) ~~,
\ee
and the cross terms give us the ${\cal O}(\cal Y)$ corrections.

Finally, there are soft mass corrections which are anarchic in flavor space that arise from non-zero VEVs for the flavor gauge boson $D$-terms. In section~\ref{sec:flavor-vevs} we saw that minimizing the $D$-term potential forced $\langle Y \rangle = \langle Y^{c} \rangle$ in 
the limit where flavon self-interactions are absent and SUSY is preserved. More generically, nonzero $\lambda_{Y},$ $\mu_{Y}$ and $m_{\cal S}$ induce small splittings,
 so the VEVs are no longer aligned and the model acquires nonzero $D$-terms 
\be
 \frac{g_{F}^{2}}{2} \left|   \tilde{Q}^{*} T^{a}_{\small{ Q}} \tilde Q  -  \tilde{\psi}_{u^{c}} T^{a}_{\small{ Q}} \tilde{\psi}_{u^{c}}^{*}     
 + Y_{u}^{*} T^{a}_{\small{ Q}} Y_{u}    -  {Y^{c}_{u}} T^{a}_{\small{ Q}} {Y^{c}_{u}}^{*}   + (u \rightarrow d) 
  \right|^{2}  + ~ SU(3)_{Q} \to  SU(3)_{U,D} ~~ .
  \label{eq:dterms-again}
\ee
There are now cross-terms that yield squark masses squared of 
order  $ \sim  g_{F}^{2}( \langle Y^{*}_{u} \rangle^{2} - \langle Y^{c\,*}_{u} \rangle^{2}) $.  To estimate the effect of flavon self interactions, we note
that (for $m_{\cal S} = 0$) the flavon VEVs depend inversely on the quartic couplings
\be
\langle Y \rangle  \simeq \frac{\!\!\!   \lambda_{S}  \, w}{         (   \lambda_{S}^{2} + \lambda_{Y}^{2}    )^{1/2}        } \approx  w \left(1 - \frac{  \lambda_{Y}^{2}}{  2 \,\lambda_{S}^{2}}\right) ~~ ,
\label{eq:trilinear-deviation}
\ee  
where $\lambda_{S}$ is the Yukawa coupling to the singlet defined in Eq.~(\ref{eq:ws}), and $\lambda_Y$ is the flavon trilinear coupling defined in Eq.~(\ref{eq:y-superpotential}). We note that $\lambda_{S}$ is identical for $Y$ and $Y^{c}$ potentials, but 
 $\lambda_{Y}$ and $\lambda_{Y^{c}}$ typically differ by small amounts, so we have done an expansion in small $\lambda_Y$.  Thus, the 
 $D$-term squark  mass squared is of order
\be
  g_{F}^{2}( \langle Y^{*}_{u} \rangle^{2} - \langle Y^{c\,*}_{u} \rangle^{2}) \sim  g_{F}^{2} \frac{ \lambda^{4}_{Y}}{2\lambda_{S}^{4}} \langle Y \rangle^{2}~~.
\label{eq:dterm-cancel}
\ee
When hidden sector soft masses are included in the flavon potential, they also upset the $\langle Y \rangle, \langle Y^{c}\rangle$ cancellation by
  shifting the VEVs by small, 
 uncorrelated amounts of order the  soft-mass scale.
 The difference in Eq\,(\ref{eq:dterm-cancel}) now becomes 
\be
  g_{F}^{2}( \langle Y^{*}_{u}\rangle^{2} - \langle Y^{c\,*}_{u}\rangle^{2}) \to   g_{F}^{2} \left(  m_{\cal S}^{2}     +      \frac{ \lambda^{4}_{Y}}{2\lambda_{S}^{4}} \langle Y\rangle^{2} \right) ~~,
  \label{eq:dterms-cancel-full}
\ee
where we have taken the typical flavon soft mass to be of order $m_{\cal S}$. 

 Summarizing, the full squark soft mass squared for $\tilde Q, \tilde{\bar{U}}$ and $\tilde{\bar{D}}$ has the form
\be
\hspace{0cm} m_{\cal S}^{2}  \biggl\{1\!
+  \!\left((c_1-c_2) + \frac{v^{2}}{m^{2}_{\cal S}} \right)  {{\cal Y}}^{\dagger}{\cal Y}  
 +      {\cal Y} \left[   {\cal O}\left( {\cal Y} \lambda_{Y} \right)  
 +      {\cal O}\left( {\cal Y}\frac{  \mu_{Y}  }{\langle Y\rangle}  \right) 
\! +    {\cal O}\left( {\cal Y}\frac{ m_{\cal S}^{2}}{\langle Y\rangle^{2}}  \right)  \right]&   \! \nonumber\\
+ g_{F}^{2} \left[ {\cal O}(1)  +     {\cal O} \left(           \frac{ \lambda^{4}_{Y}    }{\lambda_{S}^{4}     }  \frac{\langle Y\rangle^{2} }{     m_{\cal S}^{2}     } \right)     
\right]&   \!\!\!\!\biggr\} ~,~
    \label{eq:QQsoftmass}
\ee
where the first term is flavor universal and the terms proportional to $ {{\cal Y}}^{\dagger}{\cal Y}  $ are also MFV. 
We will estimate the constraints on the remaining terms in section~\ref{sec:exp}. Again, for our parametric estimates, 
we have set ${\cal A}_{\cal S} = m_{\cal S}$ for simplicity. Note that for the $\tilde Q^{*} \tilde Q$ soft mass, the $(c_1-c_2)$ term is absent, but there are additional corrections of order $v^{2} / \langle Y\rangle^{2}$  
from the tree level exchange of $\Psi$ with two Higgs insertions.  These corrections are always suppressed relative to the $m_{\cal S}^{2}/\langle Y\rangle^{2}$ terms in Eq.~(\ref{eq:QQsoftmass}), so we will not consider them further.

%%%%%%%%%%%%%%%%%%%%%%%%%%%%%%%%%%%%%%%%%%%%%%%%%%%%%%%%%%%%%%%%%%%%%%%%%%
%%%%%%%%%%%%%%%%%%%%%%%%%%%%%%%%%%%%%%%%%%%%%%%%%%%%%%%%%%%%%%%%%%%%%%%%%%
%%%%%%%%%%%%%%%%%%%%%%%%%%%%%%%%%%%%%%%%%%%%%%%%%%%%%%%%%%%%%%%%%%%%%%%%%%
%%%%%%%%%%%%%%%%%%%%%%%%%%%%%%%%%%%%%%%%%%%%%%%%%%%%%%%%%%%%%%%%%%%%%%%%%%
%%%%%%%%%%%%%%%%%%%%%%%%%%%%%%%%%%%%%%%%%%%%%%%%%%%%%%%%%%%%%%%%%%%%%%%%%%
%%%%%%%%%%%%%%%%%%%%%%%%%%%%%%%%%%%%%%%%%%%%%%%%%%%%%%%%%%%%%%%%%%%%%%%%%%
%
%													4. Experimental Constraints
%
%%%%%%%%%%%%%%%%%%%%%%%%%%%%%%%%%%%%%%%%%%%%%%%%%%%%%%%%%%%%%%%%%%%%%%%%%%
%%%%%%%%%%%%%%%%%%%%%%%%%%%%%%%%%%%%%%%%%%%%%%%%%%%%%%%%%%%%%%%%%%%%%%%%%%
%%%%%%%%%%%%%%%%%%%%%%%%%%%%%%%%%%%%%%%%%%%%%%%%%%%%%%%%%%%%%%%%%%%%%%%%%%
%%%%%%%%%%%%%%%%%%%%%%%%%%%%%%%%%%%%%%%%%%%%%%%%%%%%%%%%%%%%%%%%%%%%%%%%%%
%%%%%%%%%%%%%%%%%%%%%%%%%%%%%%%%%%%%%%%%%%%%%%%%%%%%%%%%%%%%%%%%%%%%%%%%%%

\section{Experimental Constraints}
\label{sec:exp}

In this section we consider the experimental bounds on this model coming from both direct and indirect constraints. Many of the constraints, especially the indirect ones, are unchanged from the analyses in~\cite{Csaki:2011ge} and~\cite{Grinstein:2010ve} and apply similarly to the model presented here. A more complete analysis of the low energy flavor parameters arising from the flavor gauge bosons and the exotics is given in~\cite{Buras:2011wi}.

%%%%%%%%%%%%%%%%%%%%%%%%%%%%%%%%%%%%%%%%%%%%%%%%%%%%%%%%%%%%%%%%%%%%%%%%%%
%%%%%%%%%%%%%%%%%%%%%%%%%%%%%%%%%%%%%%%%%%%%%%%%%%%%%%%%%%%%%%%%%%%%%%%%%%

%													4.1 Direct Constraints

%%%%%%%%%%%%%%%%%%%%%%%%%%%%%%%%%%%%%%%%%%%%%%%%%%%%%%%%%%%%%%%%%%%%%%%%%%
%%%%%%%%%%%%%%%%%%%%%%%%%%%%%%%%%%%%%%%%%%%%%%%%%%%%%%%%%%%%%%%%%%%%%%%%%%

\subsection{Direct Constraints}

%%%%%%%%%%%%%%%%%%%%%%%%%%%%%%%%%%%%%%%%%%%%%%%%%%%%%%%%%%%%%%%%%%%%%%%%%%
%													4.1.1 RPV bounds 
%%%%%%%%%%%%%%%%%%%%%%%%%%%%%%%%%%%%%%%%%%%%%%%%%%%%%%%%%%%%%%%%%%%%%%%%%%

\subsubsection{\small RPV Limits}
Recent LHC searches have dramatically reduced the parameter space for natural RPV spectra with
sparticle masses below the $\sim \TeV$ range. However, most of this sensitivity is driven by 
signals from more dangerous lepton violating operators $QL\bar D, LL\bar E,$ and  $L H_{u} $. Scenarios with
light, ``natural'' stops are still viable for masses above $\sim$100 GeV provided they decay primarily to dijets
 through the trilinear RPV operator $\bar U\bar D\bar D$  \cite{Evans:2012bf}. LHC trijet searches have also
recently excluded gluinos below $\sim 650$ GeV \cite{ATLAS:2012dp} by simple jet-counting searches, but
the general features of our scenario are experimentally viable for natural, sub TeV choices
of sparticle masses.

%%%%%%%%%%%%%%%%%%%%%%%%%%%%%%%%%%%%%%%%%%%%%%%%%%%%%%%%%%%%%%%%%%%%%%%%%%
%													4.1.2 Gauge Boson Production
%%%%%%%%%%%%%%%%%%%%%%%%%%%%%%%%%%%%%%%%%%%%%%%%%%%%%%%%%%%%%%%%%%%%%%%%%%

\subsubsection{\small Flavor Gauge Bosons}
\label{sec:directprod}
The inverted hierarchy mechanism ensures that most of the gauge bosons are extremely heavy; their masses are proportional to inverse Yukawa couplings. Gauge bosons that 
that mediate third generation interactions can be considerably lighter, but are constrained by limits on dijet production. Adding SUSY to the model in 
\cite{Grinstein:2010ve} does not affect the relevant signals, so the same considerations apply in our case, but the experimental limits on have improved since 2010. 
Current LHC limits \cite{CMSdijet,ATLASdijet} exclude $Z^{\prime}$ and $W^{\prime}$ states which decay to dijets below $1.5$ and $2.1$ TeV, respectively assuming SM gauge couplings.  
In section~\ref{sec:flavorful-bounds} we show that constraints on $D$-term flavor violation 
conservatively require the gauge couplings to be $g\lsim 0.03$, so naively the bound is weaker in our case, but
each of our gauge groups can supply a light gauge boson, so the effective gauge coupling is roughly
of standard model size.  This sets the strongest bound on the lightest
gauge boson masses. 

Supersymmetrization means that there will also be gauginos. These are approximately degenerate with the gauge bosons because $m_S / \langle Y \rangle \ll 1$. Since $R$-parity is approximately conserved up to small baryon number violation, gauginos must be pair produced at the LHC and will be much more difficult to observe than the gauge bosons.

%%%%%%%%%%%%%%%%%%%%%%%%%%%%%%%%%%%%%%%%%%%%%%%%%%%%%%%%%%%%%%%%%%%%%%%%%%
%													4.1.3 Exotic Production
%%%%%%%%%%%%%%%%%%%%%%%%%%%%%%%%%%%%%%%%%%%%%%%%%%%%%%%%%%%%%%%%%%%%%%%%%%

\subsubsection{\small Exotic $\Psi$ States}

The lightest exotic fermions will most often decay to $Wq$, $Zq$, or $hq$, however, the flavor violating gauge boson couplings are mixing-angle suppressed, 
so as long as the $hq$ mode is kinematically accessible, it will be the dominant decay channel. 
 Signals from 
 $\Psi \to hq\to b\bar b q$ decays are observable in trijet resonance searches \cite{ATLAS:2012dp}, which set lower bounds on RPV gluino masses of order $\sim 650$ GeV, however,
  the total rate for triplet fermions is reduced by a smaller color factor. Although a dedicated search is necessary to set limits on this 
 scenario, we expect the current bound to be of order a few hundred GeV, which is automatically satisfied if we demand  $\lambda^{\prime}_{u,d} \sim {\cal O}(1)$ from Eq.~(\ref{eq:superpotential})
 and $\langle Y_{u,d}\rangle \gsim {\cal O} (10\,\TeV)$ to be large enough for the flavor
 gauge bosons to satisfy direct production bounds. For $\lambda^{\prime}_{u,d} \sim 10^{-3} - 10^{-2}$, these states may be observable at the LHC and can give rise to 
 striking $pp \to \Psi \Psi \to (hb)(h\bar b), (ht)(h\bar t) \to (bb\bar b)(b\bar b\bar b), (b \bar b t)(b \bar b\bar t)$ signatures. The scalar partners $\tilde \Psi$
will still be roughly degenerate, but will be more difficult to produce because scalar cross sections are smaller than fermions. 
The scalars will decay down to MSSM superpartners and eventually the LSP which will decay via the RPV operator.

%%%%%%%%%%%%%%%%%%%%%%%%%%%%%%%%%%%%%%%%%%%%%%%%%%%%%%%%%%%%%%%%%%%%%%%%%%
%%%%%%%%%%%%%%%%%%%%%%%%%%%%%%%%%%%%%%%%%%%%%%%%%%%%%%%%%%%%%%%%%%%%%%%%%%

%													4.2 Indirect Constraints 

%%%%%%%%%%%%%%%%%%%%%%%%%%%%%%%%%%%%%%%%%%%%%%%%%%%%%%%%%%%%%%%%%%%%%%%%%%
%%%%%%%%%%%%%%%%%%%%%%%%%%%%%%%%%%%%%%%%%%%%%%%%%%%%%%%%%%%%%%%%%%%%%%%%%%

\subsection{Indirect Constraints}
\label{sec:exp-indirect}

%%%%%%%%%%%%%%%%%%%%%%%%%%%%%%%%%%%%%%%%%%%%%%%%%%%%%%%%%%%%%%%%%%%%%%%%%%
%													4.2.1 	four-fermi operators 
%%%%%%%%%%%%%%%%%%%%%%%%%%%%%%%%%%%%%%%%%%%%%%%%%%%%%%%%%%%%%%%%%%%%%%%%%%

\subsubsection{\small Gauge Boson Exchange}
\label{sec:4fermi}
Unlike MFV proper, the gauged flavor model allows tree-level FCNCs through flavor gauge-boson exchange even in the absence of Yukawa couplings. In the 
$M_{u,d} \to 0$ limit, the $\bar u$ fields no longer mix with the exotic fermions through the $M_{u} \psi_{u} \bar u$ coupling and every element of ${\cal V}^{u,d}_{\alpha \beta^{\prime}}$
in Eq. (\ref{eq:Yukawas}) becomes zero, but the four-Fermi operators that induce FCNCs persist: 
\be
\frac{g^{2}_{F}}{m_{V}^{2}} (\bar f \gamma^{\mu} f)(\bar f\gamma_{\mu} f) \sim \frac{1}{\langle Y \rangle^{2}} (\bar f \gamma^{\mu} f)(\bar f \gamma_{\mu} f)  ~~ , 
\label{eq:4fermi}
\ee 
where the gauge boson masses are given by $m_{V} \sim g_{F}\langle Y \rangle$, $g_{F}$ is a typical flavor gauge coupling, and $f$ is a SM fermion field whose flavor indices 
have been omitted.  Since our model inherits the same gauge interactions as \cite{Grinstein:2010ve}, 
the experimental constraints on these operators are identical.

 Note that the gauge coupling 
 cancels in Eq.\,(\ref{eq:4fermi}) and all the suppression comes from the scale of the appropriate flavon VEV. For transitions involving light 
 generation quarks,  the strongest bounds require  $\langle Y\rangle_{11, 12} \gsim 10^{3} - 10^{4} \, \TeV$ \cite{Nir:2010jr}. The corresponding bounds on 
  third generation VEVs are much weaker,  but these will be strongly constrained in section~\ref{sec:flavorful-bounds} where limits on $g_{F}$ are combined 
  with the direct production bound on the lightest gauge boson $m_{V} \sim g_{F} \, \langle Y\rangle_{33}  $ from section~\ref{sec:directprod}.

%%%%%%%%%%%%%%%%%%%%%%%%%%%%%%%%%%%%%%%%%%%%%%%%%%%%%%%%%%%%%%%%%%%%%%%%%%
%													4.2.2 	Fermion mixing
%%%%%%%%%%%%%%%%%%%%%%%%%%%%%%%%%%%%%%%%%%%%%%%%%%%%%%%%%%%%%%%%%%%%%%%%%%

\subsubsection{\small Mixing of Exotic Fermions  }
\label{sec:exotic-mixing}

%\subsubsection{\small Hadronic $Z$ Width $\boldsymbol{R_{b}}$  }
%\label{sec:Rb}
Since the exotic states $\psi$ mix with MSSM quark fields, there are new currents that couple to SM gauge bosons and modify
the theoretical prediction of $R_{b} \equiv \Gamma(Z\rightarrow b\bar b) / \Gamma(Z\rightarrow all)$, the branching ratio of the $Z$ to $b$-quarks. Since our
modifications to the model do not affect these bounds relative to the non-supersymmetric case, we can read off the relevant 
constraint from Fig.~1 of  \cite{Grinstein:2010ve}. For ${\cal O}(1)$ values of $\lambda_{d}$ and $M_{d} \gsim {\rm few} \, 100 \, \GeV$ the 
bounds are easily satisfied when the lightest down-type VEV is $\langle Y_{d}\rangle \gsim \TeV$, which is automatic given the constraints
on $\langle Y_{d}\rangle$ from $Z^{\prime}$ and $W^{\prime}$ searches -- see Sec.~\ref{sec:directprod}. 

%\subsubsection{\small CKM Element $\boldsymbol{V_{tb}}$ }
The largest mixing is among the top-like quarks and exotics, and there is an additional constraint from deviations from the CKM matrix element $V_{tb}$. The effective CKM matrix depends on the mixing angles that relate the mass/interaction eigenstates. Unitarity of the CKM is, therefore, only valid up to corrections that depend on the masses of the heavy, new particles.  Like the $R_{b}$ constraints, this observable is unaffected 
by supersymmetrization, so the bounds are identical to those summarized in Fig.~2 of \cite{Grinstein:2010ve} which shows a large, viable region of parameter
space in the $\lambda_{u}, M_{u}$ plane. In particular, for ${\cal O}(1)$ choices of $\lambda_{u}$ and $M_{u} \gsim {\rm few} \,100 \, \GeV$, the bounds are
easily satisfied as long as the smallest VEVs $\langle Y_{u,d} \rangle$ are in our range of interest $\gsim$ few $\TeV$.

%%%%%%%%%%%%%%%%%%%%%%%%%%%%%%%%%%%%%%%%%%%%%%%%%%%%%%%%%%%%%%%%%%%%%%%%%%
%													4.2.4  non-minimal flavor violation 
%%%%%%%%%%%%%%%%%%%%%%%%%%%%%%%%%%%%%%%%%%%%%%%%%%%%%%%%%%%%%%%%%%%%%%%%%%

\subsubsection{\small Flavor Violation in the MSSM Sector}
\label{sec:flavorful-bounds}

Since the exotic fermions and bosons have non-degenerate masses due to SUSY breaking and flavon self interactions, this splitting gives rise to soft-mass corrections that deviate from  MFV structure;
in full generality, the model realizes the ``flavorful supersymmetry'' scenario \cite{Nomura:2007ap}. However
in the weakly gauged, high flavor-scale limit, the deviations away from MFV are under theoretical control. 

In order to estimate the bounds on low energy observables we use the mass insertion method~\cite{Hall:1985dx}, where we define $\delta$ as the ratio of the off diagonal to the diagonal entries in the scalar mass squared matrix in the basis where the fermion masses are diagonal.  For pure MFV, $\delta_{\alpha\beta} = 0$ for $\alpha\neq \beta$. We then use the bounds on the mass insertion parameters from~\cite{Gabbiani:1996hi, Masiero:2005ua} compiled in \cite{Nomura:2007ap,Nomura:2008gg}. 

The most stringent constraints in flavorful supersymmetry come from chirality changing operators such as $\cal A$-terms,
 shown in Fig.~\ref{fig:a-term}.  We can parametrize the non-MFV contributions to $\cal A$-terms from Eq.\,(\ref{eq:a-full}) by defining
  \be
  \Delta_{\cal A}  \equiv  {\cal A}_{\cal S} \left[{\cal  O}\left({\cal Y} \lambda_{Y} \right)  +  {\cal  O}\left( {\cal Y} \frac{\mu_{Y}}{\langle Y\rangle} \right)   
   + {\cal  O}\left( {\cal Y} \frac{ m_{\cal S}^{2}}{\langle Y\rangle^{2}} \right) \right] ~~,
   \label{eq:a-anzatz}
\ee 
where  we take the overall coefficient ${\cal A_{S}}$ be of order the soft scale $m_{\cal S}$. The ${\cal O}(1)$ numbers in front of these contributions are irrelevant for setting order-of-magnitude bounds, so
we ignore them for simplicity and clarity.    

From \cite{Nomura:2007ap}, the strongest bounds on the {\it non}-MFV contributions to $\cal A$-terms come from various EDM measurements. The left/right mass insertion parameter in the up sector is given by
\be
(\delta^{u}_{       \alpha \beta   })_{          LR                }  \equiv
 \frac{                     v            (    \Delta_{\cal A})_{_{\alpha\beta}}                  }{             m_{\cal S}^{2}       }  
= \frac{                 (  \Delta_{\cal A})_{_{\alpha\beta}} (m_{u})_{  _{ \alpha \beta  }       }                  }{             m_{\cal S}^{2}   \, \,  ({\cal Y}_{u})_{_{\alpha\beta}}   \,         } ~~,
\label{eq:deltaLR}
\ee
where, $v$ is SM Higgs VEV, $(m_{u})_{        \alpha \beta         } \equiv \sqrt{  (m_{u})_{        \alpha}       (m_{u})_{        \beta  }      }$ comes from the fermion mass matrix, $(\Delta_{\cal A})_{\alpha\beta}$ is 
the non-MFV correction to the $\alpha\beta$ $\cal A$-term from Eq.\,(\ref{eq:a-anzatz}), 
 and $m_{\cal S}$ is the typical size of a 
squark mass. Note that the terms in $\Delta_{\cal A}$ are linear in Yukawa-couplings (but not the full matrices), so there is a cancellation
between this dependence and the $\cal Y$ in the denominator of  Eq.~(\ref{eq:deltaLR}) when $\delta^{u}_{LR}$ is written in terms of the quark 
mass matrix.   
Using the parametric dependencies of $\Delta_{\cal A}$, we can compare our order-of-magnitude predictions for $\delta^{u}_{LR}$ against 
the strongest representative bound on Higgs-squark $\cal A$-terms coming from the non-observation of a neutron EDM:
\be 
|{\rm Im}(\delta^{u}_{11})_{LR}| \lsim 4 \times 10^{-7} \left( \frac{m_{\cal S}}{600\, \GeV}\right)~~, 
\ee
which translates into bounds on the first generation flavon VEVs, $\langle Y \rangle_{11}$, as well as on $\lambda_{Y}$, and $\mu_{Y}/ \langle Y\rangle_{11}$. The index structure of $\langle Y \rangle$ that appears in Eq.~(\ref{eq:a-anzatz}) is discussed further in the appendix. 
 
Assuming ${\cal O}(1)$ complex phases and using $(m_{u})_{11} \approx 3\, \MeV$,
we find that the ${\cal O}( {\cal Y} m_{\cal S}^{2}/ \langle Y \rangle^{2})$ correction from $\Delta_{\cal A}$ yields the constraint 
$\langle Y \rangle_{11} \gsim 2$ TeV, which is 
 irrelevant since the scale of flavor VEVs associated with light generations must already exceed $\sim 10^{3} - 10^{4} \, \TeV$ from limits on 
four-Fermi operators in section~\ref{sec:4fermi}. 
The ${\cal O}({\cal Y} \lambda_{Y})$ and ${\cal O}({\cal Y} \mu_{Y} /  \langle Y \rangle)$ in $\Delta_{\cal A}$ are also constrained by the bounds on $(\delta^{u}_{11})_{LR}$. 
 Setting $m_{\cal S} = 600$ GeV, we require $\lambda_{Y}$ and  $\mu_{Y}/ \langle Y\rangle_{11} \lsim 10^{-3} $.

 We next compute the bounds from chirality conserving soft masses by defining 
 \be
 (\delta^{q}_{\alpha\beta})_{LL, RR} \equiv  \frac{ (\Delta m^{2}_{\tilde q})_{\alpha\beta} }{m_{\cal S}^{2}} ~~, 
 \label{eq:LLinsertion}
 \ee
where $\Delta m_{\tilde q}^{2}$ is any flavor violating (non MFV) squark mass associated with any of the $\tilde{Q}^{*} \tilde{Q} ,\tilde{U}^{*} \tilde{U}$, or  $\tilde{D}^{*} \tilde{D}$ operators.
From Eq\,(\ref{eq:QQsoftmass}) we can read off the parametric dependence of our chirality conserving corrections
\be 
\Delta m_{\tilde q}^{2} =
m_{\cal S}^{2}\,  \biggl\{    
 {\cal Y} \left[   {\cal O}\!\left({\cal Y} \lambda_{Y} \right)  \!
 +      {\cal O}\!\left( {\cal Y}\frac{  \mu_{Y}  }{\langle Y\rangle}  \right) 
 +      {\cal O}\!\left({\cal Y} \frac{ m_{\cal S}^{2}}{\langle Y\rangle^{2}}  \right)  \right]  
+ g_{F}^{2} \!\left[ {\cal O}(1)  +     {\cal O}\! \left(           \frac{ \lambda^{4}_{Y}    }{\lambda_{S}^{4}     }  \frac{\langle Y\rangle^{2} }{     m_{\cal S}^{2}     } \right)     \right] \!
  \biggr\}  ~,~
\label{eq:flavorfulmass}
\ee
where the flavor indices are suppressed. Although there are many terms to consider,  chirality-violating (left-right) constraints set much stronger bounds on 
$\lambda_{Y}$ and $\mu_{Y}/\langle Y\rangle$, so we need not evaluate them again. Furthermore, the overall scale of $\langle Y\rangle$ already has to be large to satisfy
FCNC constraints from gauge-boson exchange, so the only new constraints from chirality preserving interactions applies to the $\propto g_{F}^{2}$ terms in Eq.\,(\ref{eq:flavorfulmass}).

Unlike the left/right transitions, flavor violation arising from left/left and right/right processes is far less constrained. 
The strongest bounds on this ratio arise from $K \bar K$ mixing which requires \cite{Nomura:2007ap} 
\be
 |(\delta^{u}_{12})_{LL}|  = |(\delta^{u}_{12})_{RR}|  &\lsim& 10^{-3} - 10^{-2} ~~ ,   \label{eq:deltas1}     	\\
 |(\delta^{d}_{12})_{LL}|  = |(\delta^{d}_{12})_{RR}|  &\lsim& 10^{-2} - 10^{-1} ~~ , \\
\label{eq:deltas2}
 |(\delta^{d}_{12})_{LL}   (\delta^{d}_{12})_{RR}|\!\! \quad&\lsim& 10^{-3} ~~.
\label{eq:deltas3}
\ee
 Translating these into bounds on our model, we find 
\be
 |(\delta^{u,d}_{12})_{LL}| \sim  |(\delta^{u,d}_{12})_{RR}| \simeq g_{F}^{2}\left(1 + \frac{      \lambda_{Y}^{4}       }{     2   \lambda_{S}^{4}        }     \frac{      \langle  Y \rangle^{2}       }{      m_{\cal S}^{2}        }    \right) \lsim 10^{-3}~~,
\label{eq:LLRRbound}
\ee
which, for the first term implies $g_{F}\lsim 0.03$ and thereby places the strongest lower bound on the smallest flavor VEV since the lightest gauge boson masses 
satisfy $m_{V}\sim g_{F} \langle Y\rangle_{33} \gsim (1- 2) \,\TeV$ from the considerations in Sec.~\ref{sec:directprod}. We, thus, require the lightest
flavor VEVs to satisfy $\langle Y\rangle_{33} \gsim 30 \, \TeV$.  To estimate the bound on the second term in Eq.\,(\ref{eq:LLRRbound}), we set
 $m_{\cal S} = 600 \, \GeV$ and conservatively choose the largest flavor VEV\footnote{As long as the flavor symmetry is intact
when hidden sector SUSY breaking is communicated to the MSSM -- i.e. $M_{*} > {\rm max}\{\langle Y\rangle\}$ -- there is no upper bound on the largest VEV. 
We arrive at the upper value  $(\sim 10^{9}$ GeV) by saturating the lower bound on the {\it smallest } VEV $\langle Y\rangle_{33} \gsim 30 \, \TeV$ and scaling 
this value by the smallest (inverse) Yukawa coupling. } to be $ {\rm max}\{\langle Y\rangle\} = \langle Y \rangle_{11} = 10^{9}\, \GeV$,  
which imposes $\lambda_{Y}/ \lambda_{S} \lsim 10^{-4}-10^{-3} $.
 
In this section we have extracted bounds on up-type parameters because these are
the most tightly constrained in  \cite{Nomura:2007ap,Nomura:2008gg}, but slightly weaker bounds exist for the down-type sector as well. Since the SUSY breaking 
corrections considered here are parametrically identical in the down-sector, the constraints are similar.

%%%%%%%%%%%%%%%%%%%%%%%%%%%%%%%%%%%%%%%%%%%%%%%%%%%%%%%%%%%%%%%%%%%%%%%%%%
%													4.2.5  baryon violating transitions  
%%%%%%%%%%%%%%%%%%%%%%%%%%%%%%%%%%%%%%%%%%%%%%%%%%%%%%%%%%%%%%%%%%%%%%%%%%

\subsubsection{\small Baryon Number Violation}
\label{sec:bnvbounds}

%\label{sec:mfvbounds}
The primary constraints on the baryon number violating operators in Eq.~(\ref{eq:bnv}) are the $\Delta B = 2$ processes neutron-antineutron oscillations and 
dinucleon decay. In \cite{Csaki:2011ge}, it was shown that, under reasonable assumptions about $\tan \beta$ and hadronic matrix elements,
holomorphic MFV  sufficiently suppresses these processes to allow light ($\sim $ few 100 GeV) scalars at the weak scale. 
Since the relevant diagrams involve two insertions of flavor-violating ${\tilde b}^{*} \tilde d$ soft masses and two trilinear BNV vertices all exhibiting MFV structure, our 
non-MFV effects will appear as small corrections to squark soft masses, which shift the predicted oscillation timescale $\tau_{ \bar n- n}$ by  
\be
\tau_{\bar n - n}\sim (\tau_{\bar n - n})_{MFV}\left[1 + {\cal O}  \left( \delta_{LL/RR}\right) \right] ~~ ,
\ee
which arises from non-minimal flavor transitions among chirality preserving squark masses.  
The additional correction term is constrained to be $\lsim 10^{-3}$ by Eqs.\,(\ref{eq:deltas1}--\ref{eq:deltas3}) . Similar considerations apply to dinucleon decay, so the non-MFV effects do not
upset the relevant constraints on BNV processes; satisfying the bounds described previously in this section is 
sufficient to ensure that the desirable features of MFV SUSY are preserved.

%\subsubsection{\small Proton Decay to Gravitino}
While lepton number is a very good approximate symmetry and the usual proton decay process cannot occur, proton decay via $p \to K^{+} \tilde G$ can be mediated by just baryon number violating operators if the gravitino is sufficiently light. The diagrams for this process require one $\lambda^{\prime\prime}$ insertion and one 
off-diagonal soft mass insertion. For pure MFV SUSY, this imposes the rough constraint  \cite{Csaki:2011ge} 
\be
m_{3/2} \gsim (300\, \keV) \left(\frac{300\, \GeV}{m_{\cal S}} \right)^{2} \left( \frac{\tan \beta }{10 }\right)^{4} ~~ .
\ee
Since $m_{3/2} \sim F/M_{\rm Pl}$ and we can relate the SUSY breaking scale to the soft-mass and mediation scales 
$F \sim  m_{\cal S} M_*$, we place a lower bound on the mediation scale
\be 
M_* \gsim (10^{10} \,\GeV)      \left(\frac{300 \,\GeV}{m_{\cal S}} \right)^{3} \left( \frac{\tan \beta }{10 }\right)^{4} ~~ .
\ee
Therefore, considering SUSY breaking on the effective theory without flavor breaking is consistent because the mediation scale for SUSY breaking is higher than the scales of flavor breaking.

%%%%%%%%%%%%%%%%%%%%%%%%%%%%%%%%%%%%%%%%%%%%%%%%%%%%%%%%%%%%%%%%%%%%%%%%%%
%%%%%%%%%%%%%%%%%%%%%%%%%%%%%%%%%%%%%%%%%%%%%%%%%%%%%%%%%%%%%%%%%%%%%%%%%%
%%%%%%%%%%%%%%%%%%%%%%%%%%%%%%%%%%%%%%%%%%%%%%%%%%%%%%%%%%%%%%%%%%%%%%%%%%
%%%%%%%%%%%%%%%%%%%%%%%%%%%%%%%%%%%%%%%%%%%%%%%%%%%%%%%%%%%%%%%%%%%%%%%%%%
%%%%%%%%%%%%%%%%%%%%%%%%%%%%%%%%%%%%%%%%%%%%%%%%%%%%%%%%%%%%%%%%%%%%%%%%%%
%%%%%%%%%%%%%%%%%%%%%%%%%%%%%%%%%%%%%%%%%%%%%%%%%%%%%%%%%%%%%%%%%%%%%%%%%%
%
%													5.  Leptons
%
%%%%%%%%%%%%%%%%%%%%%%%%%%%%%%%%%%%%%%%%%%%%%%%%%%%%%%%%%%%%%%%%%%%%%%%%%%
%%%%%%%%%%%%%%%%%%%%%%%%%%%%%%%%%%%%%%%%%%%%%%%%%%%%%%%%%%%%%%%%%%%%%%%%%%
%%%%%%%%%%%%%%%%%%%%%%%%%%%%%%%%%%%%%%%%%%%%%%%%%%%%%%%%%%%%%%%%%%%%%%%%%%
%%%%%%%%%%%%%%%%%%%%%%%%%%%%%%%%%%%%%%%%%%%%%%%%%%%%%%%%%%%%%%%%%%%%%%%%%%
%%%%%%%%%%%%%%%%%%%%%%%%%%%%%%%%%%%%%%%%%%%%%%%%%%%%%%%%%%%%%%%%%%%%%%%%%%

\section{Lepton Sector}
\label{sec:leptons}

In the Standard Model, neutrinos are exactly massless, so fitting MFV to the known fact that neutrinos do have mass requires physics beyond the SM. Therefore, there is no unique prescription to apply MFV to the lepton sector~\cite{Cirigliano:2005ck,Davidson:2006bd}. 
Here, we take the attitude that the neutrino sector is analogous to the quark sector and imagine that the neutrinos get Dirac masses with a triplet of right handed neutrinos denoted $\bar N$, and 
we extend the model to include the gauged lepton group $SU(3)_{L} \times SU(3)_{E} \times SU(3)_{N}$ for the right and left-handed
 leptons. The field content in this sector is given in Table~\ref{tab:lepton-charges-dirac} and gives
 rise to the same dynamics as the quark sector; Yukawa couplings arise when flavor is broken by VEVs of flavon
 superfields and heavy exotic fields are integrated out.

%%%%%%%%%%%%%%%%%%%%%%%%%%%%%%%%%%%%%%%%%%%%%%%%%%%%%%%%%%%%%%%%%%%%%%%%%%
 %%%%%%%%%%%%%%%%%%%%%%%%%%%%%%%%%%%%%%%%%%%%%%%%%%%%%%%%%%%%%%%%%%%%%%%%%%

%													5.1 Dirac Neutrinos  

%%%%%%%%%%%%%%%%%%%%%%%%%%%%%%%%%%%%%%%%%%%%%%%%%%%%%%%%%%%%%%%%%%%%%%%%%%
%%%%%%%%%%%%%%%%%%%%%%%%%%%%%%%%%%%%%%%%%%%%%%%%%%%%%%%%%%%%%%%%%%%%%%%%%%

\subsection{Dirac Neutrinos}
The simplest lepton sector completely mimics the organization of section~\ref{sec:model} where the flavon VEVs are the only sources of 
flavor violation and the Yukawa couplings arise when $SU(3)_{L}\times SU(3)_{E}\times SU(3)_{N}$ is completely broken. In this sector, we use $Y_{\nu,e}$ to break the flavor group and $Y_{\nu,e}^{c}$ to cancel anomalies introduced
by the flavon fermionic partners. This differs from the usual seesaw scenario because there is no Majorana mass for the right handed neutrinos. While in the standard seesaw scenario the right handed neutrino is a gauge singlet which can get a Majorana mass, in our case the $\bar N$ field is charged under a flavor symmetry so it is natural to forbid its mass.  In Sec.~\ref{sec:seesaw} we will add another flavon which allows $\bar N$ to get a mass and explore the traditional seesaw scenario. 
 \begin{table}[t]
\centering
\begin{tabular}{|c|c|c|c|c|}
	\hline
		&	$SU(3)_L$		&	$SU(3)_{E }$		&		$SU(3)_{N}$   & $U(1)_{Y}$ \\       
		\hline
	$L$ 	&	$\mathbf 3$			&	$ \mathbf 1$			&		  \!\!\! $\mathbf 1 $                                 & $-1/2$   \\
	$\bar e$	&	$\mathbf 1$			&	$\mathbf 3$			&						 $\mathbf1$     & \!\!$+1$	\\
	$\bar N$ &    	$\mathbf 1$                         &       $\mathbf 1$                     &                                           $\mathbf 3$                        & \,\,\,$0$    \\
	\hline
	\hline
	$\psi_{e^{c}}$   &          $ \overline{\mathbf  3}$  &            $\mathbf  1$             &  $  \mathbf  1$  &   $\!\!+1$       \\
	$\psi_{N}$  &            $ \overline{\mathbf  3}$  &           $\mathbf 1 $             &    $\mathbf  1$  &   $\,\,\,0$       \\
	$\psi_{e}$ &            $ \mathbf 1$  &    $\overline{\mathbf  3}$                      &     $\mathbf 1$  &   $\!\!-1$       \\
	$\psi_{\nu}$ &            $\mathbf 1$  &    $\mathbf  1$                      &   $ \overline{ \mathbf 3}$ &   $\,\,\,0$       \\
	\hline 
	\hline
	$Y_{\nu}$ &            $\mathbf 3$  &    $\mathbf 1$                      &    $\mathbf 3$  &   \,\,\,0       \\
	$ Y^{c}_{\nu}$ &            $\overline{\mathbf  3}$  &    $\mathbf 1$                      &     $\overline{ \mathbf 3}$  & \,\,\,0       \\
	$Y_{e}$ &            $\mathbf 3$  &    $\mathbf 3 $                      &    $\mathbf 1$ &    \,\,\,0    \\
	$ Y^{c}_{e}$ &            $\overline{\mathbf  3}$  &    $\overline{\mathbf  3} $                      &   $\mathbf 1$ &   \,\,\,0    \\
					\hline
\end{tabular}
\vspace{0.3cm}
\caption{ The charge assignments for the lepton sector. }
\label{tab:lepton-charges-dirac}
\end{table}%

 For the pure Dirac case, the superpotential in this sector is analogous to Eq.~(\ref{eq:superpotential})
 \be
 W_{L} &=& \lambda_{e} H_{d} L \psi_{e^{c} }  +   \lambda^{\prime}_{e} Y_{e} \psi_{e} \psi_{e^{c}} + \, M_{e} \psi_{e} \bar e +  W_{Y_{L}} + ( d\to u, \, e \rightarrow \nu,  \, \bar e \to \bar N) 
 \label{eq:superpotential-leptons}
 \ee
 where the flavon self couplings are
 \be
  Y_{L} = \lambda_{Y_{L}} Y_{e}Y_{e}Y_{e} + \lambda_{Y_{L}^{c}} Y_{e}^{c}Y_{e}^{c}Y_{e}^{c} + \mu_{Y_{L}} Y_{e} Y_{e}^{c} + (e \to \nu) ~~ .
  \ee
Because the Majorana mass term for the right handed neutrinos $\bar N$ is  forbidden, the
neutrinos  are purely Dirac particles whose masses are set by Yukawa couplings ${\cal Y}_{\nu} \sim \lambda_{\nu} M_{\nu} /   \lambda^{\prime}_{\nu}  \langle Y_{\nu}  \rangle$.

The effective field theory description of MFV SUSY in \cite{Csaki:2011ge} allows lepton number violation in the Kahler potential even in the absence of neutrino masses.
With the appropriate Yukawa matrix insertions, the operator $LH_{d}^{*}$ is allowed by the ${\mathcal Z}_{3} \subset SU(3)_{L} \times SU(3)_{E}$, but only arises at 
at dimension eight and is strongly suppressed by the cutoff scale. Since our model is UV complete, it inherits the accidental lepton symmetry of the SM, so these
operators are suppressed by either the planck scale or some potential unification scale. In either case, these are irrelevant from the standpoint of IR flavor constraints.

%%%%%%%%%%%%%%%%%%%%%%%%%%%%%%%%%%%%%%%%%%%%%%%%%%%%%%%%%%%%%%%%%%%%%%%%%%
%												 5.1.1 Lepton Constraints 
%%%%%%%%%%%%%%%%%%%%%%%%%%%%%%%%%%%%%%%%%%%%%%%%%%%%%%%%%%%%%%%%%%%%%%%%%%

\subsubsection{Indirect  Constraints on Leptons}
It is straightforward to generalize the discussion of SUSY breaking from section~\ref{sec:susy-breaking}. Just like in the quark sector, the gauge  
symmetry gives rise to flavor-diagonal soft masses $\sim {\cal O}(m_{\cal S})$ with soft flavor violation arising primarily from induced Yukawa couplings 
that appear in soft masses and $\mathcal{A}$-terms. Additional flavor violation will arise in the soft terms after SUSY breaking from both the 
hidden sector described in section~\ref{sec:hidden-susy-breaking} and flavon self-interactions in section~\ref{sec:flavon-breaking}, 
but the deviations from MFV structure will again be under theoretical control and vanish in certain limits. 

As in section~\ref{sec:flavorful-bounds}, we adopt the mass insertion method to parameterize the size of FCNCs by defining the chirality violating 
mass insertion parameter
\be
(\delta^{e}_{\alpha \beta})_{LR} \equiv \frac{(\Delta_{\cal A})_{_{\alpha \beta}} (m_{e})_{_{\alpha \beta}} }{  \, m_{\cal S}^{2} \,\, ({\cal Y}_{e})_{_{\alpha \beta}}} ~~,
\ee
where $\alpha, \beta$ are generation indices, $\Delta_{A}$ is the leading {\it non} MFV contribution to the $\alpha \beta$ $\cal A$-term, and
 $(m_{e})_{\alpha \beta} \equiv  \sqrt{(m_{e})_{\alpha} (m_{e})_{\beta}} $ is the charged lepton mass matrix.  From \cite{Nomura:2008gg}, the strongest bounds in the lepton 
 sector come from $\mu \to e \gamma$ and $\mu \to eee$ rare decays, and $\mu \to e$ conversion in nuclei, which constrain 
\be
\frac{1}{\sqrt{2}} \sqrt{        |(\delta^{e}_{12})_{LR}|^{2}      +    |(\delta^{e}_{21})_{LR}|^{2}        } \,  \lsim \,4\times 10^{-6}  \left( \frac{m_{\cal S}}{ 200 \, \GeV} \right) ~~,
\ee
and the electron EDM, which requires
\be
|{\rm Im} (\delta^{e}_{11})_{LR}| \lsim 2 \times 10^{-7} \,\left( \frac{m_{\cal S}}{ 200 \, \GeV} \right) ~~.
\ee
The non-MFV corrections to lepton $\cal A$-terms are analogous to those from the quark sector; 
the deviation $\Delta_{\cal A}$ is parametrically identical to that of Eq.\,(\ref{eq:a-anzatz}) with quarks exchanged for leptons. 
 Assuming order-one complex phases, we can constrain the 
lepton VEVs $\langle Y_{e}\rangle_{12} \gsim 600 \, \GeV$ and $\langle Y_{e}\rangle_{11} \gsim 700 \, \GeV$,  which are 
slightly weaker than the corresponding bound for the 
quark-flavor scales given in section~\ref{sec:flavorful-bounds}. 
For $m_{\cal S} = 200 \, \GeV$, the bounds on flavon self-interaction parameters
 $\lambda_{Y_{e}}$ and $\mu_{Y_{e}}/ \langle Y_{e} \rangle_{11} \lsim 10^{-1}$ are also slightly weaker than those in the quark sector.

Turning to chirality-preserving flavor violation, we can define 
\be
(\delta^{e}_{\alpha \beta})_{LL, RR} \equiv \frac{   (\Delta m_{\tilde e}^{2})_{\alpha\beta}  }{  m_{\cal S}^{2}} ~~,
\ee
in analogy with Eq.~(\ref{eq:LLinsertion}), where we swap $\Delta m_{\tilde q}^{2}$ with $ \Delta m_{\tilde \ell}^{2}$, which represents all
 non-MFV soft-mass contributions to the $\tilde L^{*} \tilde L$ or $ \tilde E^{*} \tilde E$  operators. 
 For slepton soft masses at $m_{\cal S} = 200 \, \GeV$, bounds on $\mu \to e\gamma$ require \cite{Nomura:2007ap}
\be
|(\delta^{e}_{12})_{LL}| \lsim 10^{-4} -  10^{-3} ~~. 
\ee
Since the $\Delta m_{\tilde e}^{2}$ corrections are identical to those found in Eq.~(\ref{eq:flavorfulmass})
with lepton-sector parameters (e.g. $\lambda_{Y} \to \lambda_{Y_{e}}$), we find comparable bounds on lepton
sector gauge-couplings $g_{F} \lsim 10^{-2}$. The flavon self interaction parameters $\lambda_{Y_{e}} /\lambda_{S_{e}}$ also contribute corrections of order   
$g_{F}^{2}(\lambda_{Y_{e}}^{4} /\lambda_{S_{e}}^{4} ) \langle Y_{e} \rangle^{2}$. 
Following our approach in the quark sector, we constrain the latter ratio by choosing the largest flavor VEV $\langle Y_{e} \rangle \to  {\rm max}\left\{\langle Y_{e} \rangle \right\} \simeq \langle Y_{e} \rangle_{11} 
({\cal Y}_{\tau}  /{\cal Y}_{e})  \sim 10^{6} \, \GeV$.  We find that, for $m_{\cal S} = 200\,\GeV$, the ratio is bound by $\lambda_{Y_{e}} /\lambda_{S_{e} }\lsim 10^{-2}$.

Even though neutrino masses are 
purely Dirac in this case, the origin of the tiny neutrino Yukawa couplings is reminiscent of the seesaw mechanism as the couplings to $H_{u}$ 
depend on a ratio of dimensionful scales, $M_{\nu}/\langle Y_{\nu} \rangle$. Since these Yukawa couplings are of order $\sim 10^{-12}$, 
it may be necessary for the $\lambda^{\prime}_{\nu} M_{\nu}$ combination to be smaller than its corresponding size in the quark sector to keep 
the flavon VEVs below the benchmark mediation scale $M_{*} \sim 10^{10} \, \GeV$, though we note that our benchmark is merely a lower bound and the mediation scale can be raised.

Since three generations of Dirac neturinos contain six light states
that participate in
$SU(3)_{L}\times SU(3)_{N}$ gauge interactions, they can potentially
be in thermal equilibrium around the time of
Big Bang Nucleosynthesis (BBN) and double the successful SM prediction
of neutrino species in this epoch: $N_{\nu}(T_{BBN}) \simeq 3$
\cite{Beringer:1900zz}.
To maintain the success of the standard scenario in a gauged flavor
model with right handed Dirac neutrinos, the gauge bosons that couple
to the right handed neutrinos must be sufficiently weakly coupled. For
the case of gauge bosons with couplings similar to that of SM $SU(2)_L$,
this translate to requiring the gauge bosons be heavier than 3 TeV~\cite{Guadagnoli:2011id}.
In our case, the gauge coupling is smaller by an order of magnitude, so the
bound is negligible.

%%%%%%%%%%%%%%%%%%%%%%%%%%%%%%%%%%%%%%%%%%%%%%%%%%%%%%%%%%%%%%%%%%%%%%%%%%
%%%%%%%%%%%%%%%%%%%%%%%%%%%%%%%%%%%%%%%%%%%%%%%%%%%%%%%%%%%%%%%%%%%%%%%%%%

%													5.2 Majorana See Saw  

%%%%%%%%%%%%%%%%%%%%%%%%%%%%%%%%%%%%%%%%%%%%%%%%%%%%%%%%%%%%%%%%%%%%%%%%%%
%%%%%%%%%%%%%%%%%%%%%%%%%%%%%%%%%%%%%%%%%%%%%%%%%%%%%%%%%%%%%%%%%%%%%%%%%%

\subsection{Seesaw Scenario}
\label{sec:seesaw}
To include Majorana masses for right handed neutrinos, we have to add additional flavor-charged fields $M_{N}$ and $M_{N}^{c}$ which 
transform under $SU(3)_{N}$ as $\bar 6$ and 6, respectively. In addition to the $Y_{\nu}$ and $Y^{c}_{\nu}$ bifundamentals, $M_{N}$ and 
$M_{N}^{c}$ must also acquire holomorphic expectation values in order to properly execute the seesaw mechanism. 
 Modifying the superpotential from Eq. (\ref{eq:superpotential-leptons}), we now include 
 \be
 \Delta W_{L} = \frac{1}{2} M_{N} \bar N \bar N + \mu_{N} M_{N}M^{c}_{N} ~~ ,
\ee
 which yield Majorana masses for the right handed neutrino $\bar N$ when $M_{N}$ acquires a holomorphic VEV in the UV. 
 
 Although this scenario requires additional spurions, the seesaw mechanism loses some of its appeal since 
 the same physics determines the $M_{N}$ scale and the Yukawa coupling scale. In this framework,
 making the neutrino-Higgs couplings comparable to those of other SM fields implies  $\langle Y_{\nu}\rangle \sim \langle Y_{e}\rangle$, however, 
 the same physics that sets the Higgs-Yukawa couplings through $\langle Y\rangle$ also determines the scale of $\langle M_{N}\rangle$. Thus, it's not 
 clear that the seesaw mechanism in this context improves the naturalness of neutrino masses relative to the purely Dirac case. 
 Nonetheless, the added structure does not significantly modify the lepton story from \cite{Csaki:2011ge}, so we will not consider it further.

%%%%%%%%%%%%%%%%%%%%%%%%%%%%%%%%%%%%%%%%%%%%%%%%%%%%%%%%%%%%%%%%%%%%%%%%%%
%%%%%%%%%%%%%%%%%%%%%%%%%%%%%%%%%%%%%%%%%%%%%%%%%%%%%%%%%%%%%%%%%%%%%%%%%%
%%%%%%%%%%%%%%%%%%%%%%%%%%%%%%%%%%%%%%%%%%%%%%%%%%%%%%%%%%%%%%%%%%%%%%%%%%

%													6. Conclusion 

%%%%%%%%%%%%%%%%%%%%%%%%%%%%%%%%%%%%%%%%%%%%%%%%%%%%%%%%%%%%%%%%%%%%%%%%%%
%%%%%%%%%%%%%%%%%%%%%%%%%%%%%%%%%%%%%%%%%%%%%%%%%%%%%%%%%%%%%%%%%%%%%%%%%%
%%%%%%%%%%%%%%%%%%%%%%%%%%%%%%%%%%%%%%%%%%%%%%%%%%%%%%%%%%%%%%%%%%%%%%%%%%

\section{Conclusions}
\label{sec:conclusion}
Since the LHC has already put strong limits on models that predict multijet signals with  large missing energy, $R$-parity violating SUSY is, perhaps, the most compelling viable
solution to the hierarchy problem. In particular, baryon number violating RPV features multijet final states with little missing energy, so the dominant signals are 
are difficult to distinguish from the large QCD background at the LHC. These models are, therefore, quite poorly constrained  by current data. Unfortunately, abandoning $R$-parity makes the already-difficult SUSY flavor problem even worse. 
In the RPV scenario, both  superpotential and SUSY breaking sectors introduce dangerous flavor violating operators whose coefficients must be extremely 
small for the theory to be phenomenologically viable. Without additional structure to suppress flavor violation, this scenario requires extreme fine tuning. 

In this paper we have presented a model which approximately realizes the MFV framework described in~\cite{Csaki:2011ge}. This scenario, while only a hypothesis in~\cite{Csaki:2011ge}, provides a solution to SUSY flavor problem by forcing all flavor breaking parameters to be aligned with the Standard Model Yukawa matrices. Requiring the flavor breaking spurions to couple holomorphically ensures that the dominant RPV 
operators are baryon (not lepton) number violating, so the lightest (pair produced) superpartner decays predominantly to dijets, which are much harder to detect above the large QCD background.
 In our model, relatively light superpartners are still allowed and the LSP can be any particle in the MSSM. Since the SUSY breaking $\cal A$-terms  are proportional to Yukawa
 matrices, it is straightforward to engineer an IR spectrum whose LSP is either a stop or sbottom squark. 

The model presented here features a gauged $SU(3)_{Q}\times SU(3)_{U} \times SU(3)_{D}\times SU(3)_{L}\times SU(3)_{E} \times SU(3)_{N}$ flavor group and the minimal additional field content required for anomaly cancellation.
On one hand, it is a supersymmetrized version of the model presented in~\cite{Grinstein:2010ve}, which realizes ``flavorful SUSY''~\cite{Nomura:2007ap} whose 
flaovr violating parameters scale with individual Yukawa couplings, but  are not necessarily aligned with the full Yukawa matrices. 
The only terms that are not parametrically of Yukawa size arise from flavor $D$-terms, which induce anarchic, non-minimal flavor violation.
However, in this framework, all deviations away from MFV can me made small enough to avoid all experimental bounds. 

On the other hand, this model can be thought of as exactly MFV in a particular limit. This limit is as follows:
\begin{itemize}
\item The largest scale in the theory is the scale at which SUSY breaking is communicated to the visible sector, $M_*$.
\item The flavor symmetry is Higgsed by flavon VEVs $\langle Y \rangle$ at a scale parametrically below $M_*$.
\item The self interaction parameters of the flavons defined in Eq.~(\ref{eq:y-superpotential}) are small, namely $\mu_Y \ll \langle Y \rangle$
and $\lambda_Y \ll 1$.
\item The gauge coupling for the flavor gauge groups are small, $g_F \ll 1$. 
\item The size of soft SUSY breaking, $m_\mathcal{S},$ and the electroweak scale $v$ are parametrically below the flavor breaking scale $\langle Y \rangle$.
\end{itemize}
While this is a rather complicated limit, we view this not as a corner of parameter space that the model must realize, but instead as a tool for computation. 
We have elucidated the corrections away from MFV in more general regions of parameter space and find there is much room for them to be small. All low energy and collider constraints have been detailed in section~\ref{sec:exp} and we find that this model can be viable with weak scale MSSM superpartners. 

In this work, we have focused mostly on the quark sector, but we consider the lepton sector in section~\ref{sec:leptons}. Introducing a triplet of right handed neutrinos $\bar N$ which transform under their own $SU(3)_N$ leads to a natural realization of pure Dirac neutrinos. While the $\bar N$'s are normally gauge singlets which can get a Majorana mass, the flavor gauge symmetry forbids this Majorana mass, making pure Dirac neutrinos a more natural possibility than the standard seesaw scenario. The neutrino Yukawa couplings must be very small, but they remain technically natural because they arise from superpotential parameters. In order to implement the seesaw scenario, one must add
 additional flavons which transform as the symmetric $6$ and $\bar 6$ representations of $SU(3)_N$. We briefly consider this scenario in section~\ref{sec:seesaw}.

In addition to the full MSSM spectrum, this model predicts many
of the same features as the low-scale gauged flavor model with some 
exotic states that could be discovered at the LHC. Most of the bounds derived in \cite{Grinstein:2010ve} apply to this scenario, and we have updated the experimental 
limits to reflect the latest constraints. Current LHC $Z^{\prime}$ and $W^{\prime}$ limits require the lightest flavor gauge boson masses to be 
in the low TeV range, while fourth generation searches require the exotic fermion masses to be above a few hundred GeV. 
While this model only contains the minimum number of states consistent with supersymmetry and the new gauge symmetry, it has a large number of new states which could be accesible to the LHC in the near future.

\section*{Acknowledgments}
We thank Carlos Tamarit for helpful conversations and comments on the draft. We also thank Csaba Csaki, Roberto Franceschini, Ben Heidenreich, Natalia Toro, and Itay Yavin for helpful conversations. 
Research at the Perimeter Institute is supported in part by the Government of Canada through Industry Canada, 
and by the Province of Ontario through the Ministry of Research and Information (MRI). DS is supported in part by
the NSF under grant PHY-0910467, gratefully acknowledges support from the Maryland Center for Fundamental Physics, and is appreciative of support and hospitality from the Galileo Galilei Institute during the completion of this work.

%%%%%%%%%%%%%%%%%%%%%%%%%%%%%%%%%%%%%%%%%%%%%%%%%%%%%%%%%%%%%%%%%%%%%%%%%%%%%%%%%%%%%%%%%%%%%%
%%%%%%%%%%%%%%%%%%%%%%%%%%%%%%%%%%%%%%%%%%%%%%%%%%%%%%%%%%%%%%%%%%%%%%%%%%%%%%%%%%%%%%%%%%%%%%
%%%%%%%%%%%%%%%%%%%%%%%%%%%%%%%%%%%%%%%%%%%%%%%%%%%%%%%%%%%%%%%%%%%%%%%%%%%%%%%%%%%%%%%%%%%%%%
%
%  												Appendix 
%
%%%%%%%%%%%%%%%%%%%%%%%%%%%%%%%%%%%%%%%%%%%%%%%%%%%%%%%%%%%%%%%%%%%%%%%%%%%%%%%%%%%%%%%%%%%%%%
%%%%%%%%%%%%%%%%%%%%%%%%%%%%%%%%%%%%%%%%%%%%%%%%%%%%%%%%%%%%%%%%%%%%%%%%%%%%%%%%%%%%%%%%%%%%%%
%%%%%%%%%%%%%%%%%%%%%%%%%%%%%%%%%%%%%%%%%%%%%%%%%%%%%%%%%%%%%%%%%%%%%%%%%%%%%%%%%%%%%%%%%%%%%%

 \section*{Appendix: Exotic Mass Diagonalization }
 \renewcommand{\theequation}{A.\arabic{equation}}
\setcounter{equation}{0}

In the MFV limit the flavor scale satisfies $\langle Y\rangle \gg  v, m_{S}, \mu_{Y}$ and the flavon self-interaction couplings
 are negligible $\lambda_{Y }\ll 1$, the 
exotics acquire masses almost exclusively from the superpotential in Eq.~(\ref{eq:superpotential}), and SUSY is an approximate 
symmetry among the exotic states. In this appendix, we will diagonalize the exotic mass matrix in this limit. 
 For simplicity of exposition, we set the ${\cal O}(1)$ coefficients $\lambda_{u,d}, \lambda^{\prime}_{u,d}$ to unity throughout. 
 
 The mass terms in the scalar potential are
\be 
\sum_{\alpha} \left|\frac{\partial W}{\partial \psi_{  \bar u_\alpha}      } \right|^{2} &=&  \left|     \langle Y_{u}\rangle^{\alpha \beta^{\prime }}   \!\!  \tilde { \, \, \psi_{u} }_{\beta^{\prime}}      \right|^{2} = 
    \langle Y_{u}\rangle^{*}_{\alpha \gamma^{\prime }}     \langle Y_{u}\rangle^{\alpha \beta^{\prime }}      \tilde { \, \, \psi_{u}^{*} }^{\gamma^{\prime}  }      \tilde { \, \, \psi_{u} }_{\beta^{\prime}}       \\ 
\sum_{\beta^{\prime}} \left| \frac{\partial W}{\partial    { \bar u  }^{\beta^{\prime}}  } \right|^{2} &= &  \left|  M_{u}       \tilde { \, \, \psi_{u} }_{\beta^{\prime}}      \right|^{2} = M_{u}^{2}  \tilde { \, \, \psi_{u}^{*} }^{\beta^{\prime}} \tilde { \, \, \psi_{u} }_{\beta^{\prime}}   
\\ 
\sum_{\beta^{\prime}}      \left| \frac{\partial W}{        \partial  {\psi_{u}}_{\beta^{\prime}}     } \right|^{2} &=&    
\left|     \langle Y_{u}\rangle^{\alpha \beta^{\prime}} \!\!\!\!   \tilde { \, \, \, \psi_{u^c} }_{\alpha}     + M_{u} \tilde{   \,  {\bar u} }^{\beta^{\prime}}  \right|^{2} \nonumber \\ 
&=&  \langle Y_{u}\rangle^{*}_{\gamma \beta^{\prime}   }           \langle Y_{u}\rangle^{\alpha \beta^{\prime}} \!\!\!\!                        \tilde { \, \, \, \psi_{u^c}^{*}      }^{\gamma}                              \tilde { \, \, \, \psi_{u^c}      }_{\alpha}  +
M_{u}^{*} \langle Y_{u}\rangle^{\alpha \beta^{\prime}} \!    \tilde{   \,  {\bar u} }_{\beta^{\prime}}^{*}          \!\!    \tilde { \, \, \, \psi_{u^c}      }_{\alpha}  
 \nonumber \\ &&+ \langle Y_{u}\rangle^{*}_{\gamma \beta^{\prime}} \!   M_{u}             \!\!    \tilde { \, \, \, \psi_{u^c}^{*}      }^{\gamma}  \tilde{   \,  {\bar u} }^{\beta^{\prime}} + M_{u}^{2} \tilde{   \,  {\bar u} }_{\beta^{\prime}}^{*}  \tilde{   \,  {\bar u} }^{\beta^{\prime}}
\ee
which populate the scalar mass-squared matrix $m_{\tilde \psi}^{2}$.
In the same basis (prior to EWSB) the fermion mass terms are 
\be
\langle Y_{u} \rangle^{\alpha \beta^{\prime}}  { \psi_{u^c}}_{ \alpha}  {\psi_{u}}_{\beta^{\prime}} + M_{u}{\bar u}^{\beta^{\prime}} {\psi_{u}}_{\beta^{\prime}}  + h.c. ~~,
\ee
which give rise to the mass matrix  $M_{\psi}$.
Note that $M_{\psi}^{\dagger} M_{\psi} = m^{2}_{\tilde \psi}$, so exotic particles are mass degenerate with their superpartners; both matrices 
$M_{\psi}$ and $m^{2}_{\psi}$ are diagonalized by the 
same unitary transformations. In the EWSB preserving limit, this system has three zero eigenvalues whose eigenvectors become the MSSM $\bar U $ fields. 

For the purpose of illustration, we can perform gauge transformations to make $\langle Y_{u}\rangle$  diagonal and real so that  
$\langle Y_{u}\rangle^{\alpha \beta^{\prime}} = 
\langle Y^{*}_{u}\rangle_{\alpha \beta^{\prime}}$, which
 moves the CKM matrix over to the down-type VEVs.  
After diagonalizing the mass matrix, the massless MSSM eigenstates $\bar U^{\alpha} \equiv (\bar U^{1},\bar U^{2}, \bar U^{3})  $ 
can be written in terms of 
the corresponding flavor eigenstates  
$\psi_{u^{c}}^{\alpha}$ and $ \bar u^{ \alpha}$
\be
\bar U^{\alpha} =  \frac{           1         }{\sqrt{    1 +     \frac{M^{2}_{u} }{\, \langle Y_{u} \rangle_{\alpha \alpha}^{2} }    }}   
    \left(          \bar u^{\alpha} -   \frac{M_{u} }{\, \langle Y_{u} \rangle_{\alpha \alpha} } \, \psi_{u^c}^{\alpha}    \right)  ~~.
\ee
 Since the elements of the diagonalization matrix $\cal V$ are defined by the $\psi_{u^{c}}$ coefficients, and $\cal V \propto \cal Y$, then
  (up to order-one factors) the scale of Yukawa couplings is set by the appropriate $\sim M_{u}/\langle Y_{u}\rangle$ ratio.
   This justifies our approach in section~\ref{sec:flavorful-bounds} where the bounds on $(\delta^{u}_{11})_{LR}$ translate into a 
  lower bound on the $\langle Y_{u}\rangle_{11}$ element.

\end{document}